\documentclass[aps,pra,reprint,superscriptaddress]{revtex4-2}\usepackage{amsmath,amssymb,amsfonts,bm}
\usepackage{physics} 
\usepackage{newtxtext}

\usepackage{graphicx}
\usepackage[dvipsnames]{xcolor}
\usepackage[normalem]{ulem}

\usepackage[colorlinks=true, citecolor=blue, urlcolor=blue, linkcolor=blue]{hyperref}

\begin{document}

\title{Observable signatures of exceptional points from left–right eigenstate distinction}
\author{Leela Ganesh Chandra Lakkaraju}
\affiliation{Pitaevskii BEC Center, CNR-INO and Dipartimento di Fisica, Universit\`a di Trento, Via Sommarive 14, Trento, I-38123, Italy}	
\affiliation{INFN-TIFPA, Trento Institute for Fundamental Physics and Applications, Via Sommarive 14, Trento, I-38123, Italy}	
\author{Soumik Bandyopadhyay}
\affiliation{Pitaevskii BEC Center, CNR-INO and Dipartimento di Fisica, Universit\`a di Trento, Via Sommarive 14, Trento, I-38123, Italy}	
\affiliation{INFN-TIFPA, Trento Institute for Fundamental Physics and Applications, Via Sommarive 14, Trento, I-38123, Italy}	
\affiliation{School of Physics, Indian Institute of Science Education and Research Thiruvananthapuram, 695551, India}
\author{Sudipto Singha Roy}
\affiliation{Department of Physics, Indian Institute of Technology (ISM) Dhanbad, IN-826004, Dhanbad, India}
\author{Philipp Hauke}
\affiliation{Pitaevskii BEC Center, CNR-INO and Dipartimento di Fisica, Universit\`a di Trento, Via Sommarive 14, Trento, I-38123, Italy}	
\affiliation{INFN-TIFPA, Trento Institute for Fundamental Physics and Applications, Via Sommarive 14, Trento, I-38123, Italy}

\begin{abstract}

Non-Hermitian quantum systems exhibit qualitatively distinct physical behavior compared to Hermitian systems, a prime example being spectral singularities known as exceptional points. 
Their relevance in, e.g., quantum sensing, unidirectional transport,  and robust lasing makes it important to be able to identify exceptional points through observable features of a many-body system. Here, using as an example a one-dimensional complex $XY$ spin chain realizing both rotation-time ($\mathcal{RT}$)- and parity-time ($\mathcal{PT}$)-symmetric regimes, we develop a framework for detecting exceptional points based on the distinction between left and right eigenvectors of the Hamiltonian, which in a non-Hermitian system are no longer the adjoint of each other. 
We first show that a global measure constructed from the difference between the Hamiltonian and its adjoint locates exceptional points via distinct non-analytic behavior. At the level of observables, differences in local spin correlations evaluated on the right and left eigenstates provide a reliable static detection scheme. 
In contrast, static bipartite entanglement measures fail to capture this distinction, urging us to study the quantum dynamics of the model. Following a sudden quench, we demonstrate that the time-averaged right–left entanglement entropy difference directly encodes signatures of the exceptional point. In the $\mathcal{RT}$-symmetric regime, it exhibits a pronounced peak at the exceptional point, whereas in the $\mathcal{PT}$-symmetric regime it behaves as an order-parameter-like quantity, remaining finite in one phase and vanishing at the transition. Our results establish a direct link between the structure of non-Hermitian eigenstates and observable signatures of exceptional points, providing a practical route to identify them in existing quantum simulators.

\end{abstract}

\maketitle

\section{Introduction}

Non-Hermitian quantum physics extends beyond conventional paradigms, giving rise to phenomena without Hermitian counterparts. Two fundamental features underpin this framework. First, non-Hermitian Hamiltonians host exceptional points \cite{exceptional_point_textbook, Heiss_physics_of_exceptional_points}, spectral singularities where eigenvalues and eigenvectors coalesce, rendering the Hamiltonian defective. Second, their eigenstates form distinct right and left manifolds associated with the Hamiltonian and its adjoint. These features underlie functionalities such as unidirectional invisibility \cite{lin2011unidirectional, feng2013experimental}, coherent perfect absorption \cite{longhi2010pt, chong2011pt}, topological lasing \cite{harari2018topological, bandres2018topological}, enhanced sensitivity due to the associated Riemann surface topology \cite{hodaei2017enhanced, chen2017exceptional,agarwal2025criticalquantummetrologyusing,agarwal2026quantumenhancedsensinginterplaylongrange}, and higher power storage in quantum batteries \cite{non_herm_battery_2024}. Non-Hermitian dynamics govern phenomena including resonances \cite{moiseyev2011non}, superradiance \cite{dicke1954coherence}, the quantum Zeno effect \cite{facchi2008quantum}, measurements of the fluctuation--dissipation relation and current operators~\cite{kevin_hauke_prxq_2022,geier2021noninvasivemeasurementcurrentsanalog}, and non-Hermitian criticality \cite{ashida2017parity}. In this context, a natural question in many-body systems is how exceptional points manifest in the distinct right and left state manifolds, and whether this distinction can be exploited to detect their presence through accessible observables, both in equilibrium and out of equilibrium.

To systematically exploit these non-Hermitian features, recent work by some of the authors established a quantitative framework that connects Hamiltonian non-Hermiticity to experimentally accessible observables \cite{soumik_sudipto_hauke_2025}. The central idea is to compare the quantum evolution generated by a non-Hermitian Hamiltonian and its adjoint, which drive a given initial state toward distinct right and left quantum ensembles associated with their respective eigenstates \cite{sudipto_soumik_non_hermitian_eth_prl_2025}. Evaluating expectation values independently within these ensembles provides a direct operational probe of non-Hermiticity. A key insight of this framework is that non-Hermiticity at the level of the Hamiltonian, understood as the intrinsic non-Hermitian structure of the generator of dynamics, does not in general translate directly to physical observables. Depending on the observable and the parameter regime, expectation values in the two ensembles can either differ substantially or coincide, even when the underlying dynamics is strongly non-Hermitian. Observable differences therefore, emerge selectively and can reveal signatures of the underlying spectral structure, including the presence of exceptional points.

Building on this formalism, we investigate a generalized one-dimensional $XY$ spin chain with complex anisotropy and complex transverse magnetic field. This model provides a versatile many-body setting that naturally realizes both rotation--time ($\mathcal{RT}$)-symmetric  \cite{Song_RT_symm} and parity--time ($\mathcal{PT}$)-symmetric regimes \cite{heralded_pt_}, giving rise to qualitatively distinct classes of exceptional points and critical behavior, ranging from symmetry-breaking transitions to changes in spectral topology \cite{PT_sym_imaginary_magnetic_field,non_herm_extension_batchelor_2025}. To characterize these exceptional boundaries, we employ two complementary approaches. First, we use a trace-norm-based quantifier that captures the global degree of non-Hermiticity and develops characteristic non-analytic features at exceptional points. Second, we construct observable-based diagnostics by evaluating local spin expectation values independently in the right and left ground-state manifolds, thereby establishing a direct connection to measurable quantities.

To access the physical consequences of these spectral singularities beyond static properties, we further investigate the non-equilibrium dynamics following a sudden quantum quench into the non-Hermitian regime. We initialize the system in the exact ground state of the corresponding Hermitian model and subsequently evolve it under the non-Hermitian dynamics and its adjoint counterpart, tracking the resulting right and left trajectories. This dynamical perspective is essential, as static bipartite entanglement measures remain insensitive to the underlying non-Hermitian structure: despite the distinct nature of the right and left ground states, their entanglement properties coincide across all parameter regimes. In contrast, non-Hermitian dynamics generate a finite entanglement entropy difference between the evolutions driven by the Hamiltonian and by its adjoint.

We find that this dynamical response exhibits qualitatively different behavior in the two symmetry classes. In the $\mathcal{RT}$-symmetric regime, the time-averaged right--left entanglement entropy difference exhibits a pronounced peak at the exceptional point, identifying the boundary between complex and real spectral regions. Beyond the transition, the signal decreases but remains finite, reflecting a qualitative change in the underlying dynamics. In the $\mathcal{PT}$-symmetric regime, by contrast, the entanglement difference behaves as an order-parameter-like quantity: it remains finite in the topologically non-trivial phase and drops sharply to zero at the exceptional point, providing a direct dynamical signature of the transition. By combining Hamiltonian-level diagnostics with observable-based probes and intrinsically dynamical signatures, our results establish a coherent and experimentally accessible framework for detecting exceptional points and characterizing non-Hermitian many-body systems.

The paper is organized as follows. In Sec.~\ref{sec:model_exceptional_point}, we introduce the generalized non-Hermitian $XY$ model and define its corresponding exceptional points for both the $\mathcal{RT}$ and $\mathcal{PT}$ symmetries. Section~\ref{sec:hamiltonian_difference} details the detection of these exceptional boundaries using the Hamiltonian-based trace-norm quantifier. In Sec.~\ref{sec:observable_difference}, we construct local observable-based quantifiers and examine the static differences between the right and left manifolds. Section~\ref{sec:dynamics_entropy_difference} explores the dynamical evolution of the entanglement entropy difference following a non-Hermitian quench. Finally, we summarize our findings and conclude in Sec.~\ref{sec:conclusion}.

\section{Non-Hermitian $\mathcal{RT}$- and $\mathcal{PT}$-symmetric models and exceptional points}
\label{sec:model_exceptional_point}
\begin{figure*}
    \includegraphics[width=0.8\linewidth]{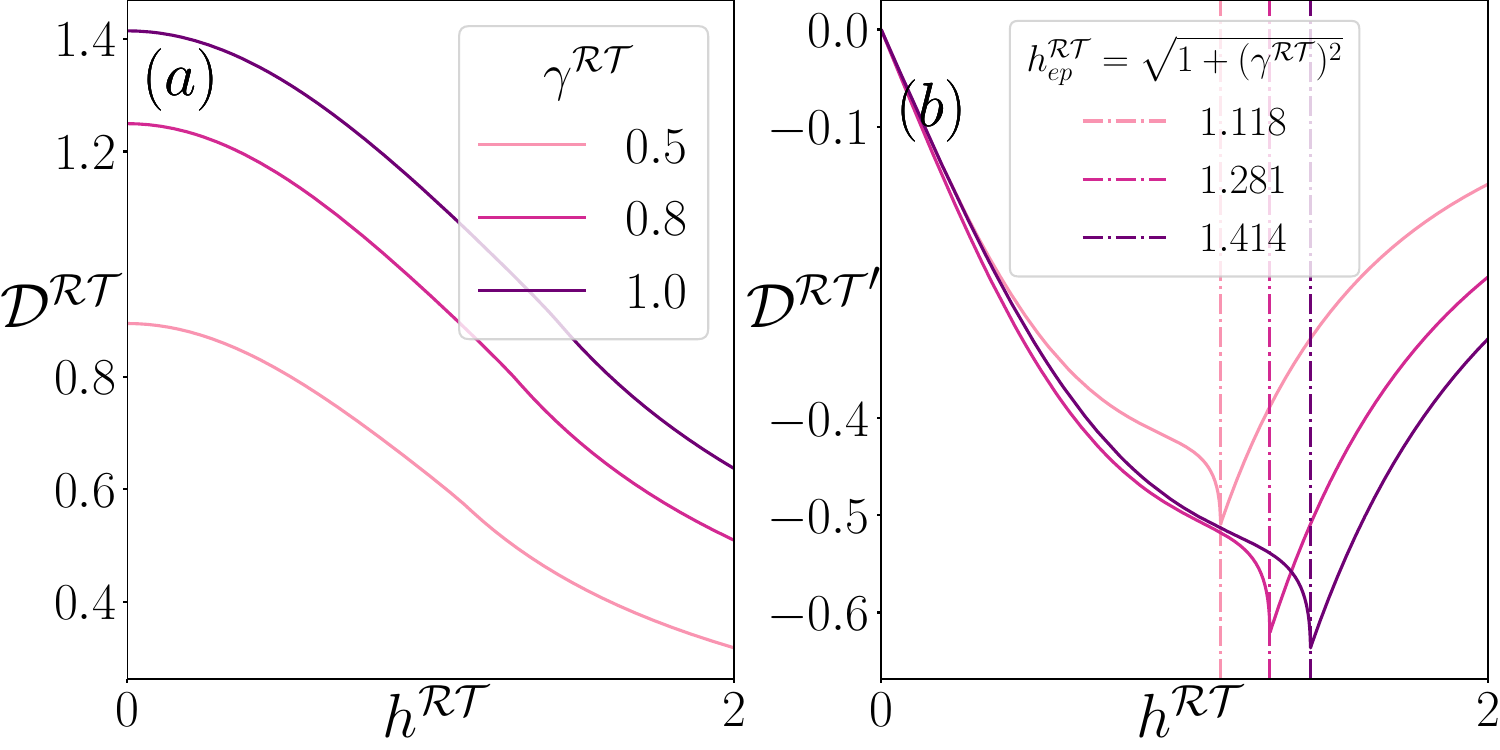}
    \caption{(a) The Hamiltonian-based quantifier $\mathcal{D^{RT}}$ defined in Eq.~(\ref{eq:ham_diff_rt}) vs.\ on-site magnetic field, for varying anisotropy strengths $\gamma^\mathcal{RT} \in \{0.5, 0.8, 1.0\}$, increasing from lighter to darker curves for the $\mathcal{RT}$-symmetric model $H^\mathcal{RT}$.  (b) The derivative $\mathcal{D^{RT}}^\prime =  d \mathcal{D^{RT}}/dh^{\mathcal{RT}}$ shows non-analytic behavior locating the exceptional point $h^{\mathcal{RT}}_{\mathrm{ep}}$ (indicated by dash-dotted vertical lines). Data for system size $N = 10000$. Both axes are dimensionless.}
    \label{fig:rt_sym_hamiltonian_difference}
\end{figure*}

Within the broader landscape of non-Hermitian many-body physics, the one-dimensional complex $XY$ spin chain provides a paradigmatic setting for exceptional-point phenomena, characterized by a transition from real to complex eigenvalues in the rotation -time ($\mathcal{RT}$)-symmetric case \cite{Song_RT_symm}, and by changes in spectral topology in parity-time ($\mathcal{PT}$)-symmetric systems \cite{PT_sym_imaginary_magnetic_field}. To systematically investigate these phenomena, we consider a generalized lattice Hamiltonian characterized by a complex anisotropy parameter $\tilde{\gamma}$ and a complex transverse magnetic field $\tilde{h}$ \cite{non_herm_extension_batchelor_2025}. Assuming uniform ferromagnetic interactions, the unified Hamiltonian takes the form
\begin{equation}
    \tilde{H} = -\sum_{r=1}^{N} \Big[ \frac{1+\tilde{\gamma}}{4}\sigma_r^x\sigma_{r+1}^x + \frac{1-\tilde{\gamma}}{4}\sigma_r^y\sigma_{r+1}^y \Big ] - \frac{\tilde{h}}{2} \sum_{r=1}^N \sigma_r^z,
\label{eq:ham_general}
\end{equation}
where $\sigma^\alpha$ ($\alpha = x,y,z$) denote the standard Pauli matrices, $\tilde{\gamma} \in \mathbb{C}$ dictates the complex spin-exchange anisotropy, and $\tilde{h} \in \mathbb{C}$ represents the normalized complex on-site magnetic field. Depending on the chosen parameter regime, this generalized form encapsulates two fundamentally distinct classes of non-Hermitian symmetries, each harboring unique behavior. 

The first regime constitutes the \textbf{$\mathcal{RT}$-symmetric $XY$ model} ($H^{\mathcal{RT}}$). This occurs when the anisotropy is strictly imaginary ($\tilde{\gamma} = i\gamma^{\mathcal{RT}}$, with $\gamma^{\mathcal{RT}} \in \mathbb{R}$) and the magnetic field remains real ($\tilde{h} = h^{\mathcal{RT}} \in \mathbb{R}$) \cite{Song_RT_symm}. The symmetry of this model is the invariance under the combined operation of time-reversal $\mathcal{T}$ (which enacts complex conjugation, $i \to -i$) and a global spin rotation $\mathcal{R} = \prod_r \exp(-i\frac{\pi}{4}\sigma^z_r)$, which maps $\sigma^x \to \sigma^y$ and $\sigma^y \to -\sigma^x$. An effective non-Hermitian Hamiltonian of this form arises when two system spins interact with a single bath spin \cite{recog_keshav_tanoy_aditi} (see Appendix~\ref{sec:app_rt_derivation} for details).

The second regime constitutes the \textbf{$\mathcal{PT}$-symmetric $XY$ model} ($H^{\mathcal{PT}}$). In this case, the nature of the parameters is reversed: the anisotropy is strictly real ($\tilde{\gamma} = \gamma^{\mathcal{PT}} \in \mathbb{R}$), while the transverse magnetic field is purely imaginary ($\tilde{h} = ih^{\mathcal{PT}}$, with $h^{\mathcal{PT}} \in \mathbb{R}$) \cite{PT_sym_imaginary_magnetic_field}. In this framework, the parity operator $\mathcal{P}$ acts as a global spatial inversion in spin space equivalent to a $\pi$-rotation around the $x$-axis ($\mathcal{P} = \prod_r \sigma_r^x$) mapping $\sigma^x \to \sigma^x$ and $\sigma^{y,z} \to -\sigma^{y,z}$. When combined with the time-reversal operator $\mathcal{T}$, the $\mathcal{PT}$ operation preserves the dissipative imaginary field and ensures global Hamiltonian invariance. For a three-level atom, an effective Hamiltonian $H^{\mathcal{PT}}$ is obtained by postselecting trajectories conditioned on no decay events \cite{heralded_pt_,PT_sym_imaginary_magnetic_field} (see Appendix~\ref{app:pt_sym_derivation} for details).

Despite its non-Hermiticity, this many-body system remains analytically tractable through a sequence of exact mappings. We first map the localized spin operators to non-local spinless fermions via the Jordan--Wigner transformation \cite{jordan-wigner}, defined as $c_r = (\prod_{j<r} \sigma_j^z) \sigma_r^-$, where $\sigma_r^- = (\sigma_r^x - i\sigma_r^y)/2$. Restricting our analysis to the even-parity sector, we impose anti-periodic boundary conditions on the fermionic fields ($c_{N+1} = -c_1$). Applying a discrete Fourier transform, $c_r = \frac{1}{\sqrt{N}} e^{-i\pi/4} \sum_{p} e^{i \phi_p r} c_p$ with quantized momenta $\phi_p = \frac{\pi}{N}(2p-1)$, where $p = 1,2, \ldots \frac{N}{2}$, allows us to decouple the system into independent two-level momentum sectors. In the Nambu spinor basis $\Psi_p^\dagger = (c_p^\dagger, c_{-p})$, the Hamiltonian is block-diagonalized as $\tilde{H} = \bigoplus_{\phi_p > 0} \tilde{H}_{\phi_p}$, with each $2 \times 2$ sector taking the form \cite{bm1,ising_scipost_santoro}
\begin{equation}  
\tilde{H}_{\phi_p}=
\begin{bmatrix}
-\tilde{h}-\cos \phi_p & -\tilde{\gamma} \sin \phi_p \\
\tilde{\gamma} \sin \phi_p & \tilde{h}+\cos \phi_p
\end{bmatrix}.
\label{eq:ham_p}
\end{equation}

The diagonalization of this non-Hermitian block matrix requires a non-unitary Bogoliubov transformation \cite{heralded_pt_,Song_RT_symm}. By introducing a generally complex Bogoliubov angle $\theta_p \in \mathbb{C}$, constrained by the relation $\tan \theta_p = \frac{ \tilde{\gamma} \sin \phi_p}{\tilde{h} + \cos \phi_p}$, we obtain the complex dispersion relation $E_p = \pm \sqrt{(\tilde{h} + \cos \phi_p)^2 + \tilde{\gamma}^2 \sin^2 \phi_p}$. 

Because $\tilde{H} \neq \tilde{H}^\dagger$, the right and left eigenbases of the system do in general not coincide. The right eigenstates satisfy
\[
\tilde{H} \, | \Psi^R \rangle = E_n \, | \Psi^R \rangle,
\]
while the left eigenstates are defined through
\[
\tilde{H}^\dagger \, | \Psi^L \rangle = E_n^* \, | \Psi^L \rangle.
\] Formed by the vacuum of the Bogoliubov quasiparticles, the corresponding right and left many-body ground states adopt a BCS-like structure
\begin{align}
    |\Psi^R\rangle &= \prod_{\phi_p > 0} \left( \cos \frac{\theta_p}{2} + i \sin \frac{\theta_p}{2} c_p^\dagger c_{-p}^\dagger \right) |0\rangle \text{ and } \label{eq:right_gs}  \\
    |\Psi^L\rangle &= \prod_{\phi_p > 0} \left( \cos \frac{\theta_p^*}{2} + i \sin \frac{\theta_p^*}{2} c_p^\dagger c_{-p}^\dagger \right) |0\rangle, \label{eq:left_gs}
\end{align}
where $|0\rangle$ denotes the Bogoliubov vacuum. The physical significance of this complex spectrum becomes most apparent at the exceptional point (EP)---a purely non-Hermitian defect where both the eigenvalues and their corresponding eigenvectors completely coalesce \cite{exceptional_point_textbook,bender_ropp_2007,bender_prl_1998}. 

\begin{figure*}
\includegraphics[width=\linewidth]{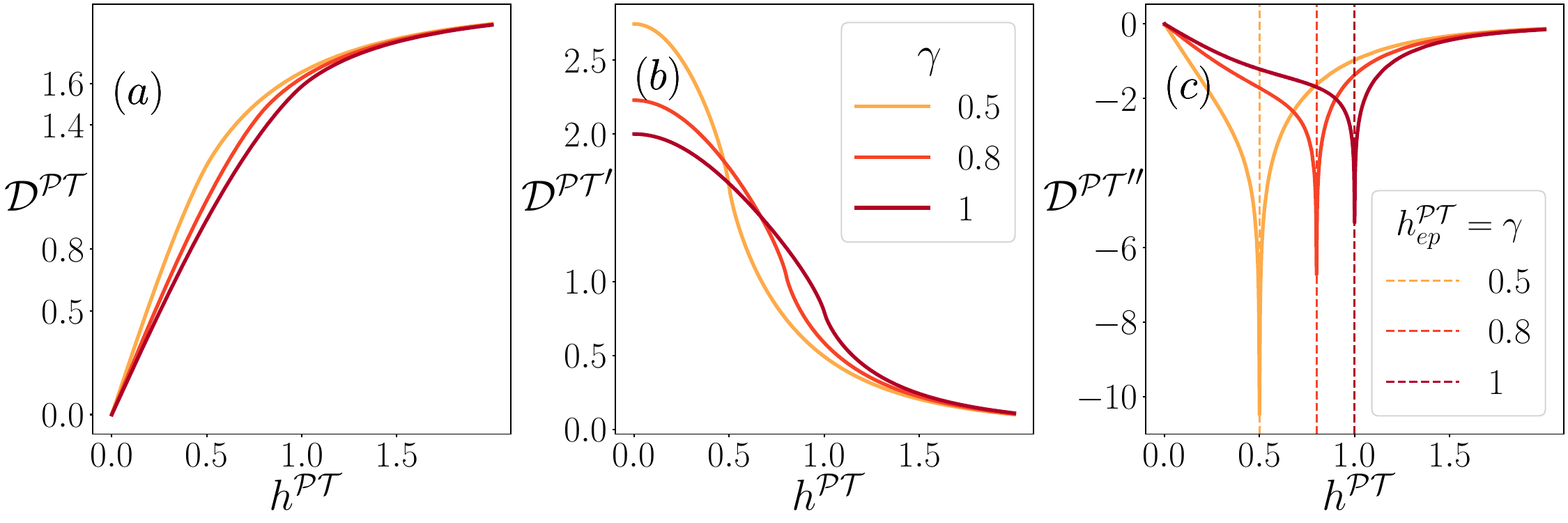}
    \caption{(a) The Hamiltonian-based quantifier $\mathcal{D^{PT}}$ defined in Eq.~(\ref{eq:ham_diff_pt}) vs. on-site magnetic field,  for  anisotropy parameter $\gamma^{\mathcal{PT}} \in \{0.5, 0.8, 1.0\}$ increasing from light to dark shades. (b) The derivative $\mathcal{D^{PT}}^\prime = d \mathcal{D^{PT}}/dh^{\mathcal{PT}}$. (c) The second derivative $\mathcal{D^{PT}}^{\prime \prime} = d^2 \mathcal{D^{PT}}/d{h^{\mathcal{PT}}}^2$. 
    While $\mathcal{D}^{\mathcal{PT}}$ and its first derivative are smooth throughout parameter space, $\mathcal{D^{PT}}^{\prime \prime}$ displays non-analytic behavior at the EP located at $h^{\mathcal{PT}}_{\mathrm{ep}}$
    (indicated by the dash-dotted vertical lines). All other specifications are same as in Fig.~\ref{fig:rt_sym_hamiltonian_difference}.}
    \label{fig:pt_sym_hamiltonian_difference}
\end{figure*}
Let us first consider the $\mathcal{RT}$-symmetric regime. In this model, the exceptional point emerges exactly at the magnetic field $h^{\mathcal{RT}}_\mathrm{ep} = \sqrt{1+(\gamma^{\mathcal{RT}})^2}$ \cite{Song_RT_symm}, acting as the strict boundary for spontaneous symmetry breaking. When the magnetic field is tuned above this threshold ($h^{\mathcal{RT}} > h^{\mathcal{RT}}_{ep}$), the system resides in the unbroken $\mathcal{RT}$-symmetric phase, characterized by a completely real many-body energy spectrum. Conversely, as the field is lowered across the EP ($h^{\mathcal{RT}} < h^{\mathcal{RT}}_{ep}$), the $\mathcal{RT}$ symmetry is spontaneously broken, causing the real eigenvalues to sharply bifurcate into complex conjugate pairs.

The $\mathcal{PT}$-symmetric regime presents a fundamentally different behavior. Because the transverse field is natively dissipative (purely imaginary), the dispersion relation is intrinsically complex across the entire parameter space. Consequently, the EP, which appears at $h^{\mathcal{PT}}_{ep} = \gamma^{\mathcal{PT}}$, does not herald a real-to-complex spectral transition, but rather signifies a topological phase transition \cite{non_herm_extension_batchelor_2025}, separating two distinct phases characterized by a non-Hermitian winding number $w$. Following standard non-Hermitian topological band theory, this invariant measures the accumulated phase winding across the Brillouin zone, defined continuously as $w = \frac{1}{2\pi} \int_{-\pi}^{\pi} d\phi_p \partial_{\phi_p} \Phi(\phi_p)$, where $\Phi(\phi_p) = \arctan\left( \frac{ \gamma^{\mathcal{PT}} \sin \phi_p}{ih^\mathcal{PT} + \cos \phi_p} \right)$. Note that even though $\Phi$ is imaginary, $w$ is always real \cite{non_hermitian_winding_number_chaohong_2020}. For magnetic strengths below the threshold ($h^{\mathcal{PT}} < h^{\mathcal{PT}}_{ep}$), the system is driven into a non-trivial topological phase ($|w| = 1$), which is intimately linked to dynamical topological observables \cite{non_hermitian_winding_number_chaohong_2020} and robust area-law scaling of entanglement. Conversely, tuning the imaginary field past the EP ($h^{\mathcal{PT}} > h^{\mathcal{PT}}_{ep}$) forces the winding number to zero, rendering the system topologically trivial.

\section{Detection of Exceptional Points Using a Hamiltonian Quantifier}
\label{sec:hamiltonian_difference}

Exploiting the distinct algebraic structure of non-Hermitian systems, some of us recently proposed a quantifier to measure the degree of non-Hermiticity \cite{soumik_sudipto_hauke_2025}. This Hamiltonian-based quantifier is defined as  
\begin{equation}
    \mathcal{D} = \frac{||\tilde{H}-\tilde{H}^\dagger||_1}{||\tilde{H}||_1} = \frac{\int_0^\pi d \phi_p||\tilde{H}_{\phi_p}-\tilde{H}_{\phi_p}^\dagger||_1 }{\int_0^\pi d \phi_p||\tilde{H}_{\phi_p}||_1},
    \label{eq:quantifier_def}
\end{equation}
where $||A||_1 $ denotes the Schatten-1 norm (trace norm), representing the sum of the singular values of matrix $A$, and where in the second equality we used the analytically known momentum-space decomposition as given in Eq.~\eqref{eq:ham_p}.  
Remarkably, in the family of models considered in this paper, the behavior of $\mathcal{D}$ and its derivatives with respect to the tuning parameters reveals the location of the exceptional points (EPs). However, the specific analytic signatures depend fundamentally on the system's underlying symmetry class.

\subsection{Signatures in the $\mathcal{RT}$-Symmetric Phase}

For $H^\mathcal{RT}$, the quantifier $\mathcal{D}$ admits an analytic evaluation upon exploiting the momentum-space decomposition introduced in Eq.~\eqref{eq:ham_p}, leading to
\begin{equation}
    \mathcal{D}^{\mathcal{RT}} = \frac {\int_0^\pi d \phi_p4 \gamma^\mathcal{RT}\sin \phi_p} {\int_0^\pi d \phi_p|a^\mathcal{RT}| + |b^\mathcal{RT}|} = \frac {8 \gamma^\mathcal{RT}} {\int_0^\pi d \phi_p|a^\mathcal{RT}| + |b^\mathcal{RT}|},
    \label{eq:ham_diff_rt}
\end{equation}
where $a^\mathcal{RT} = h^\mathcal{RT} + \cos \phi_p + \gamma^\mathcal{RT}\sin \phi_p$ and $b^\mathcal{RT} = h^\mathcal{RT} + \cos \phi_p - \gamma^\mathcal{RT}\sin \phi_p$. 

Figure~\ref{fig:rt_sym_hamiltonian_difference}(a) shows $\mathcal{D}^{\mathcal{RT}}$ as a function of the real on-site magnetic field $h^{\mathcal{RT}}$, for varying anisotropy strengths $\gamma^\mathcal{RT} \in \{0.5, 0.8, 1.0\}$. 
As the magnetic field increases, $\mathcal{D}^{\mathcal{RT}}$ exhibits a smooth, monotonic decrease across the parameter space, reflecting the gradual suppression of the relative non-Hermiticity by the real magnetic field. However, the signature of the exceptional point is encoded in the curvature of this decay. The behavior of the quantifier can be separated into two distinct analytical regimes separated by the exceptional point:

(i) \textit{Broken $\mathcal{RT}$ phase ($h^{\mathcal{RT}} < h^{\mathcal{RT}}_{ep}$):} In this non-Hermitian regime, $\mathcal{D}^{\mathcal{RT}}$ yields a concave dependence on $h^{\mathcal{RT}}$. The system's complex spectrum determines the decay rate, which accelerates as the field strength increases.

(ii) \textit{Unbroken $\mathcal{RT}$ phase ($h^{\mathcal{RT}} \ge h^{\mathcal{RT}}_{ep}$):} Once the system crosses the exceptional point into the regime of purely real eigenvalues, the quantifier transitions to a strictly convex tail that asymptotically approaches zero.
\begin{figure*}
    \centering
    \includegraphics[width=\linewidth]{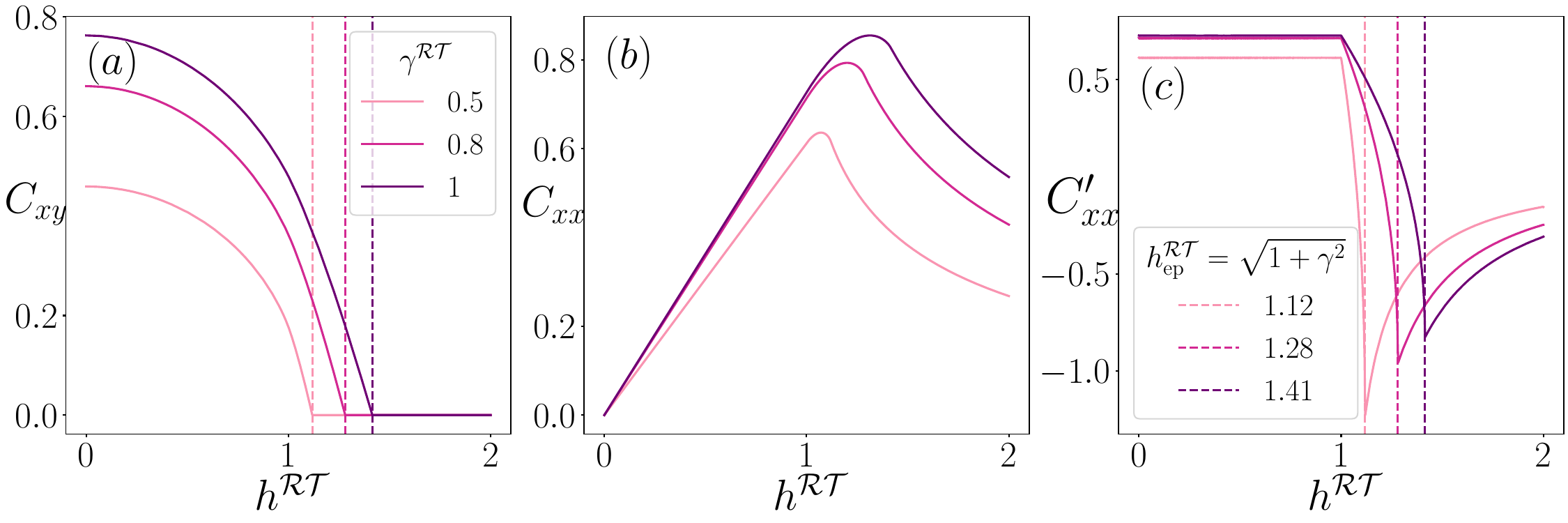}
    \caption{Observable-based quantifiers defined in Eq.~(15) in the $\mathcal{RT}$-symmetric regime. (a) Off-diagonal correlator difference $C_{xy}$, (b) diagonal correlator difference $C_{xx}$, and (c) its derivative $C_{xx}^\prime = dC_{xx}/dh^{\mathcal{RT}}$ as functions of the transverse magnetic field $h^{\mathcal{RT}}$. The anisotropy parameter varies as $\gamma^{\mathcal{RT}} \in \{0.5, 0.8, 1.0\}$ from lighter to darker curves. Vertical dashed lines indicate the corresponding exceptional point positions $h^{\mathcal{RT}}_{\mathrm{ep}} = \sqrt{1+({\gamma^{\mathcal{RT}}})^2}$ for each $\gamma^{\mathcal{RT}}$. The off-diagonal correlator $C_{xy}$ remains non-zero in the broken phase and vanishes in the unbroken phase, thus acting as an order-parameter, whereas the diagonal correlator $C_{xx}$ changes curvature at $h_{\mathrm{ep}}^{\mathcal{RT}}$ as indicated by the non-analyticity in $C_
    {xx}^\prime$. All quantities are dimensionless.}
    \label{fig:rt_sym_observable_difference}
\end{figure*}
This inversion of curvature at the boundary between the two phases translates into a stark non-analyticity in the first derivative. As illustrated in Fig.~\ref{fig:rt_sym_hamiltonian_difference}(b), the magnitude of the derivative ${\mathcal{D}^{\mathcal{RT}}}^\prime = d\mathcal{D}^{\mathcal{RT}}/dh^{\mathcal{RT}}$ grows steadily until it reaches a sharp, non-differentiable cusp. This cusp occurs, within numerical precision, at the exceptional point $
    h_{\mathrm{ep}}^\mathcal{RT} = \sqrt{1+(\gamma^\mathcal{RT})^2}$,
marked by the vertical dash-dotted lines. Immediately beyond $h_{\mathrm{ep}}^\mathcal{RT}$, the magnitude of the derivative abruptly decays. This pronounced cusp in the first derivative provides a definitive and easily discernible signature of the non-Hermitian $\mathcal{RT}$ symmetry-breaking transition, cleanly isolating it from standard Hermitian critical phenomena. 
\subsection{Signatures in the $\mathcal{PT}$-Symmetric Phase}
\begin{figure*}
    \centering
    \includegraphics[width=0.8\linewidth]{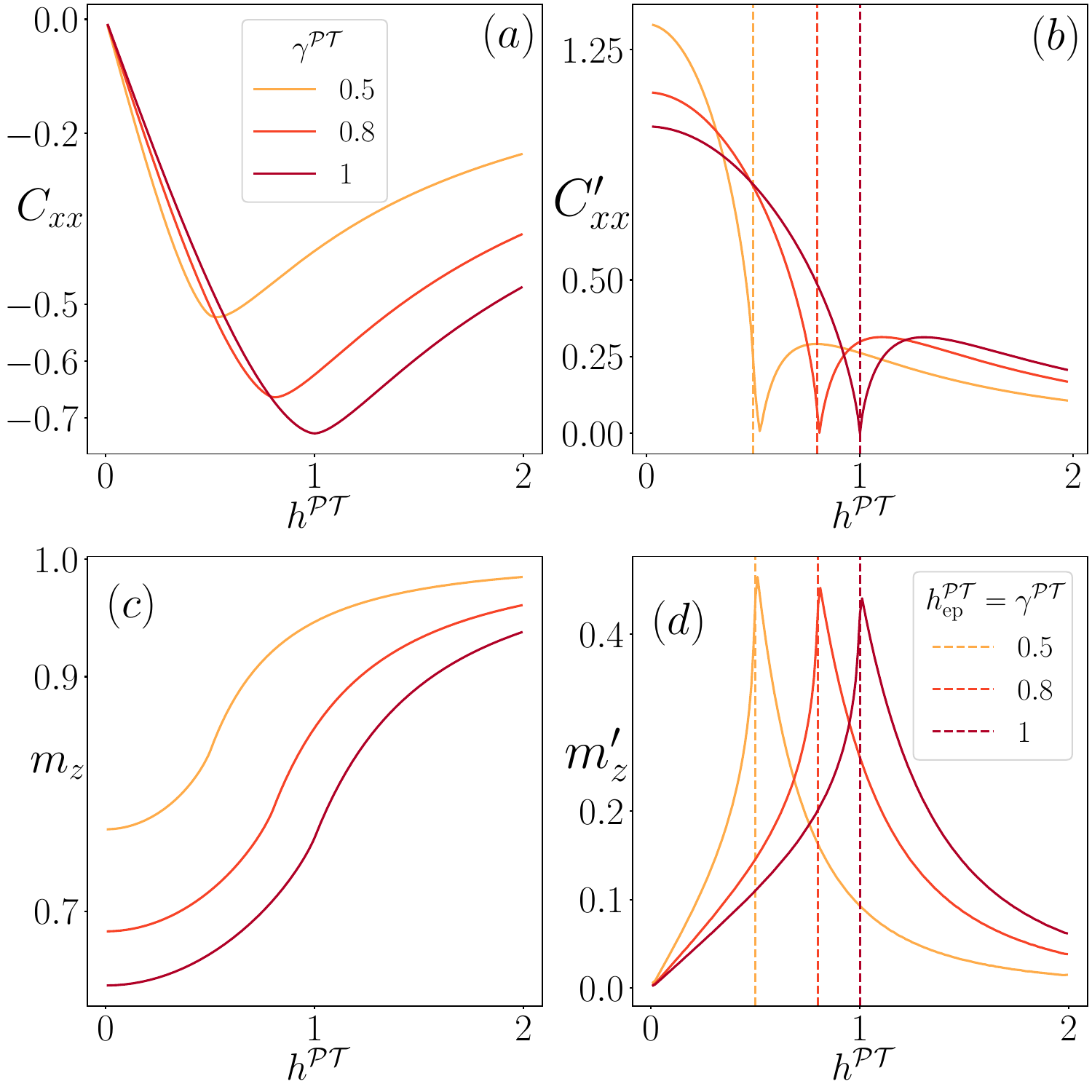}
    \caption{Observable-based quantifiers defined in Eq.~(15) in the $\mathcal{PT}$-symmetric regime. (a) Diagonal correlator difference $C_{xx}$, (b) its derivative  $C_{xx}^{\prime}=dC_{xx}/dh^{\mathcal{PT}}$, (c) magnetization difference $m_z$, and (d) its derivative $m_{z}^{\prime}=dm_z/dh^{\mathcal{PT}}$, shown as functions of the imaginary transverse field $h^{\mathcal{PT}}$. The anisotropy parameter varies as $\gamma^{\mathcal{PT}} \in \{0.5, 0.8, 1.0\}$ from lighter to darker curves. While $C_{xx}$ and $m_z$ remain smooth across the transition, their derivatives display pronounced non-analytic behavior at the exceptional point. Vertical dashed lines indicate the corresponding exceptional point positions $h^{\mathcal{PT}}_{\mathrm{ep}} = \gamma^{\mathcal{PT}}$. All quantities are dimensionless.}
    \label{fig:pt_sym_observable_difference}
\end{figure*}
The framework must be adapted when analyzing the $\mathcal{PT}$-symmetric $XY$ model, where the anisotropy $\gamma^\mathcal{PT}$ is real and the non-Hermiticity is driven by an imaginary magnetic field, $ ih^\mathcal{PT}$. Applying Eq.~(\ref{eq:quantifier_def}) to this regime yields the quantifier
\begin{equation}
    \mathcal{D}^{\mathcal{PT}} = \frac {\int_0^\pi  d\phi_p  4 |h^\mathcal{PT}|} {\int_0^\pi d \phi_p ~ a^{\mathcal{PT}}  + b ^ {\mathcal{PT}}} = \frac{4\pi |h^\mathcal{PT}|}{\int_0^\pi d \phi_p ~ a^{\mathcal{PT}}  + b ^ {\mathcal{PT}}},
    \label{eq:ham_diff_pt}
\end{equation}
where $a^\mathcal{PT} = \sqrt{\cos^2\phi_p + (h^\mathcal{PT}+\gamma^\mathcal{PT}\sin\phi_p)^2}$ and $b^\mathcal{PT} = \sqrt{\cos^2\phi_p + (h^\mathcal{PT}-\gamma^\mathcal{PT}\sin \phi_p)^2}$. The Hamiltonian-quantifier $\mathcal{D}^{\mathcal{PT}}$ begins at zero for the Hermitian limit ($h^\mathcal{PT} = 0$) and increases monotonically with the field strength, as illustrated in the first panel of Fig.~\ref{fig:pt_sym_hamiltonian_difference}(a). The first derivative ${\mathcal{D}^{\mathcal{PT}}}^{\prime } = d\mathcal{D}^{\mathcal{PT}}/dh^\mathcal{PT}$ remains continuous and smooth across the exceptional point (see Fig.~\ref{fig:pt_sym_hamiltonian_difference} (b)) in contrast to the $\mathcal{RT}$ transition. 

Instead, the exceptional signature is pushed to a higher order. Evaluating the \textit{second derivative} ${\mathcal{D}^{\mathcal{PT}}}^{\prime \prime}= d^2\mathcal{D}^{\mathcal{PT}}/d{h^{\mathcal{PT}}}^2$ uncovers a sharp divergence (a non-analytic dip), as shown in  Fig.~\ref{fig:pt_sym_hamiltonian_difference}(c). This divergence occurs at $h^\mathcal{PT} = \gamma^{\mathcal{PT}}$, i.e., it coincides with the exceptional point for the $\mathcal{PT}$-symmetric model. 

In summary, the Hamiltonian quantifier $\mathcal{D}$ offers a unified mechanism for detecting non-Hermitian exceptional points. While the $\mathcal{RT}$-breaking transition manifests as a non-analyticity in the first derivative of the Hamiltonian quantifier, the $\mathcal{PT}$-breaking transition is indicated by a divergence in its second derivative.

\section{Observable-Based Quantification of the Exceptional Point}
\label{sec:observable_difference}

To systematically characterize the non-Hermitian phase transitions without requiring access to the full complex many-body spectrum, we construct local observable-based quantifiers. Departing from the conventional biorthogonal metric \cite{brody_iop_2014} typically introduced in non-Hermitian systems to define expectation values, we instead evaluate observables directly with respect to the independent right and left ground-state manifolds \cite{turkeshi_prb_2023, turkeshi_schrio_stochastic_resetting_2022}. This choice is natural in non-Hermitian settings, where non-unitary evolution under the Hamiltonian and its adjoint prepares distinct right and left ensembles \cite{soumik_sudipto_hauke_2025,sudipto_soumik_non_hermitian_eth_prl_2025}. In this non-Hermitian framework, the ground state is defined as the minimum-energy state when the spectrum is purely real, and as the eigenstate with the largest imaginary part of the eigenvalue—corresponding to the dominant long-time (steady) state—when the spectrum becomes complex \cite{ground_state_zhang_2022,ground_state_schiro_scipost_2025}. For a generic operator $\mathcal{O}$, the expectation value within the state $\nu \in \{R, L\}$ is defined via the normalized Dirac inner product:
\begin{equation}
    \langle \mathcal{O} \rangle_{\nu} = \frac{\langle \Psi^\nu | \mathcal{O} | \Psi^\nu \rangle}{\langle \Psi^\nu | \Psi^\nu \rangle}.
\end{equation}

We investigate the longitudinal magnetization $\mathcal{M}_z^\nu = \frac{1}{N}\sum_r \langle \sigma_r^z \rangle_\nu$ and the nearest-neighbor spin correlation functions $\mathcal{C}_{\alpha\beta}^\nu = \frac{1}{N}\sum_r \langle \sigma_r^\alpha \sigma_{r+1}^\beta \rangle_\nu$ for $\alpha, \beta \in \{x, y, z\}$. To compute these quantities, the spin operators are mapped to the Bogoliubov quasiparticle basis using the transformations outlined in Sec.~\ref{sec:model_exceptional_point}. Because the states $|\Psi^R\rangle$ and $|\Psi^L\rangle$ are vacua of free fermions, they are inherently Gaussian. Consequently, the expectation values of multi-spin operators reduce to combinations of two-point fermionic contractions via Wick's theorem.

The state $|\Psi^\nu\rangle$ is fully parameterized by the complex amplitudes $u_p^\nu$ and $v_p^\nu$. Following the canonical Bogoliubov transformation \cite{heralded_pt_,Song_RT_symm} that diagonalizes the block matrix in Eq.~(\ref{eq:ham_p}), these amplitudes are identified as $u_p^R = \cos(\theta_p/2)$ and $v_p^R = \sin(\theta_p/2)$ for the right state, and $u_p^L = \cos(\theta_p^*/2)$ and $v_p^L = \sin(\theta_p^*/2)$ for the left state. The particle number distribution $n^\nu(\phi_p)$ and the anomalous pairing amplitude $f^\nu(\phi_p)$ are given by
\begin{equation}
    n^\nu(\phi_p) = \frac{|v_p^\nu|^2}{|u_p^\nu|^2 + |v_p^\nu|^2}, \quad f^\nu(\phi_p) = \frac{(u_p^\nu)^* v_p^\nu}{|u_p^\nu|^2 + |v_p^\nu|^2}.
\end{equation}

By substituting the inverse Fourier transforms of the fermionic operators into the definitions of the spin observables, the expressions decouple into sums over these momentum-space contractions. In the thermodynamic limit ($N \to \infty$), the discrete summation over the quantized momenta $\phi_p$ transitions to a continuous integral over the positive half of the Brillouin zone, $\phi \in [0, \pi]$. The longitudinal magnetization evaluates to
\begin{equation}
    \mathcal{M}_z^\nu = 1 - \frac{2}{\pi} \int_0^\pi d\phi \, n^\nu(\phi).
\end{equation}

For the nearest-neighbor correlators, the Wick expansion generates specific combinations of the normal and anomalous terms. The correlation functions evaluate to
\begin{align}
    \mathcal{C}_{xx}^\nu &= \frac{1}{\pi} \int_0^\pi d\phi \Big[ \cos \phi \left(2 n^\nu(\phi) - 1\right) - 2 \sin \phi \, \text{Re}[f^\nu(\phi)] \Big], \\
    \mathcal{C}_{yy}^\nu &= \frac{1}{\pi} \int_0^\pi d\phi \Big[ \cos \phi \left(2 n^\nu(\phi) - 1\right) + 2 \sin \phi \, \text{Re}[f^\nu(\phi)] \Big], \\
    \mathcal{C}_{xy}^\nu &= \frac{2}{\pi} \int_0^\pi d\phi \, \sin \phi \, \text{Im}[f^\nu(\phi)].
\end{align}
The longitudinal correlator $\mathcal{C}_{zz}^\nu$ can be written compactly by defining the intermediate integrals $G^\nu = \frac{1}{\pi} \int_0^\pi d\phi \cos\phi \, n^\nu(\phi)$ and $S^\nu = \frac{1}{\pi} \int_0^\pi d\phi \sin\phi \, f^\nu(\phi)$, yielding
\begin{equation}
    \mathcal{C}_{zz}^\nu = (\mathcal{M}_z^\nu)^2 - 4 (G^\nu)^2 - 4 |S^\nu|^2.
\end{equation}

Applying this analytical framework, the primary quantity of interest in this work for detecting exceptional points is the difference between the expectation values evaluated in the right and left ground-state manifolds. We define these observable differences as \begin{equation}
    C_{\alpha\beta} \equiv \mathcal{C}_{\alpha\beta}^R - \mathcal{C}_{\alpha\beta}^L \quad \text{and} \quad m_z \equiv \mathcal{M}_z^R - \mathcal{M}_z^L.
    \label{eq:ob_quant}
\end{equation} 
By monitoring these differences across the parameter space, we identify signatures of the phase transitions for both symmetry classes.  

In the $\mathcal{RT}$-symmetric regime (Fig.~\ref{fig:rt_sym_observable_difference}), the exceptional point at $h_{\mathrm{ep}}^{\mathcal{RT}} = \sqrt{1+(\gamma^{\mathcal{RT}})^2}$ demarcates the boundary between the broken and unbroken symmetry phases. The difference in the magnetization vanishes entirely ($m_z = 0$) across all regimes. Instead, the phase transition is captured by the bipartite correlation functions. Most notably, the off-diagonal correlator difference $C_{xy}$ acts effectively as an order parameter for the non-Hermitian transition as seen in Fig.~\ref{fig:rt_sym_observable_difference}(a). It maintains a finite value in the broken-symmetry phase ($h^{\mathcal{RT}} < h_{\mathrm{ep}}^{\mathcal{RT}}$) and drops to zero upon entering the unbroken phase at $h^{\mathcal{RT}} \geq h_{\mathrm{ep}}^{\mathcal{RT}}$. In this unbroken regime, even though the Hamiltonian is explicitly non-Hermitian, the expectation values of the right and left states align, so that the observable differences vanish, making some of the system's observable behavior similar to that of a Hermitian model. Furthermore, the transition imprints a clear signature on the diagonal correlator difference $C_{xx}$. While $C_{xx}$ varies continuously and changes curvature at the exceptional point, its first derivative with respect to the magnetic field, $dC^\mathcal{RT}_{xx}/dh^{\mathcal{RT}}$, exhibits a sharp non-analyticity at $h_{\mathrm{ep}}^{\mathcal{RT}}$ (see Figs.~\ref{fig:rt_sym_observable_difference} (b) and (c)).

As mentioned above, in the $\mathcal{PT}$-symmetric regime the exceptional point resides at $h_{\mathrm{ep}}^{\mathcal{PT}} = \gamma^{\mathcal{PT}}$ and identifies a topological phase transition. In this case, essentially all the observable differences exhibit signatures of the transition (Fig.~\ref{fig:pt_sym_observable_difference}). While both the diagonal correlator difference $C_{xx}$ (Fig.~\ref{fig:pt_sym_observable_difference}(a))  and the magnetization difference $m_z$ (Fig.~\ref{fig:pt_sym_observable_difference}(c)) are smooth continuous functions of the field strength $h^{\mathcal{PT}}$, the location of the exceptional point is resolved by their corresponding derivatives with respect to the field. The derivative of the correlator difference, $dC_{xx}/dh^{\mathcal{PT}}$, develops a non-analytic cusp at the exceptional point (Fig.~\ref{fig:pt_sym_observable_difference}(b)). Simultaneously, the derivative of the magnetization difference, $dm_z/dh^{\mathcal{PT}}$, displays a pronounced divergence at $h_{\mathrm{ep}}^{\mathcal{PT}}$ (see Fig.~\ref{fig:pt_sym_observable_difference}(d)), providing a signature of the underlying topological transition point or the exceptional point. 

As these local spin observables and their differences successfully capture the non-Hermitian critical boundaries, it is natural to inquire whether quantum information measures exhibit similar features. Using the spatial correlation observables derived above, one can construct the reduced density matrix for nearest-neighbour spins to evaluate bipartite entanglement properties. However, von Neumann entanglement entropy, logarithmic negativity, and quantum mutual information yield identical values for the right and left ground state across all parameter regimes considered. %The difference between the independent states strictly vanishes for these equilibrium information-theoretic metrics in the lowest-energy configuration. 
One may conjecture that this symmetric behavior is restricted to the ground state, and that higher excited states may indeed encode resolvable signatures of the left--right asymmetry. To probe this question, in the next section, we investigate non-equilibrium protocols. Indeed, the distinct time evolutions generated by the Hamiltonian and its adjoint probe different parts of the spectrum, leading to pronounced differences in the dynamically generated entanglement.

\section{Difference in entanglement entropy of dynamical right and left vectors}
\label{sec:dynamics_entropy_difference}
\begin{figure*}
    \includegraphics[width=\linewidth]{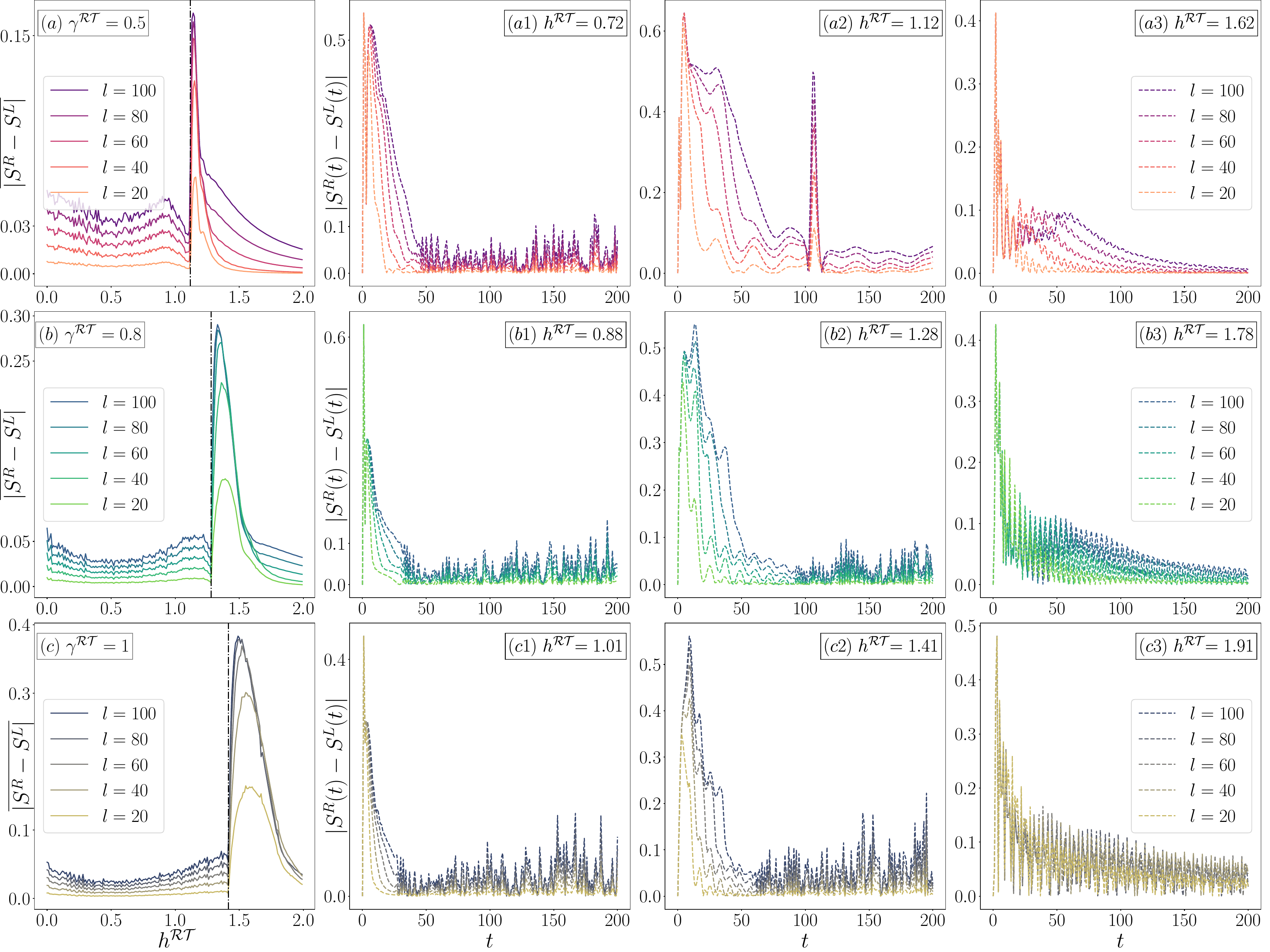}
    \caption{
Time-averaged entanglement entropy difference $\overline{|S_l^R - S_l^L|}$ following a sudden quench from the ground state of a Hermitian $XY$ model ($\gamma=0.5$, $h=0.5$) into a non-Hermitian $\mathcal{RT}$-symmetric $XY$ model with imaginary anisotropy. Panels in the left column [(a)–(c)] show the time-averaged quantity as a function of the transverse magnetic field $h^{\mathcal{RT}}$ for subsystem sizes $l \in \{20,40,60,80,100\}$. The dash-dotted vertical lines mark the corresponding exceptional points $h^{\mathcal{RT}}_{\mathrm{ep}} = \sqrt{1+({\gamma^{\mathcal{RT}}})^2}$. The remaining panels [(a1)–(c3)] display the real-time evolution $|S_l^R(t) - S_l^L(t)|$ up to $t=200$ for representative values of $h^{\mathcal{RT}}$ indicated in each panel. For each row, the three panels correspond to field values below [(a1), (b1), (c1)], at [(a2), (b2), (c2)], and above [(a3), (b3), (c3)] the exceptional point. Rows correspond to anisotropy values $\gamma^{\mathcal{RT}} = 0.5$ [(a)], $0.8$ [(b)], and $1$ [(c)]. Colors indicate subsystem sizes $l$, from lighter to darker shades corresponding to smaller to larger $l$. The time-averaged entanglement difference develops a well-defined peak in the close vicinity of the exceptional point across all values of $\gamma^{\mathcal{RT}}$.  Here, $N = 10000.$ All quantities are dimensionless.
}
    \label{fig:rt_sym_entro_difference}
\end{figure*}
As quantum information measures show no difference in the left and right ground states, one may naturally wonder if the same holds in the non-equilibrium dynamics of the non-Hermitian system. Further, from an experimental standpoint, realizing quench dynamics is often considerably more feasible on near-term quantum simulators than achieving convergence to the ground state, which can pose significant challenges even in the Hermitian regime \cite{gorshkov_ground_state_hard_prl_2019}. 

Initializing a system in a well-defined and simple Hermitian ground state and subsequently evolving it under an effective non-Hermitian Hamiltonian via a sudden quench provides a direct, executable protocol in modern quantum simulation platforms using localized dissipation \cite{pt_dynamics_experiment_photonics,pt_symmetry_dynamics_nitrogen_vacancy,pt_symmetry_dynamics_photonics_topological_edge, pt_symmetry_dynamics_photons,pt_symmetry_experiment_floquet_joglekar,pt_symmetry_dynamics_superconducting,sudipto_soumik_non_hermitian_eth_prl_2025,kevin_hauke_prxq_2022,matteo_hauke_non_herm}. 
Here, we initialize the system in the state $|\Psi_0\rangle$ defined as the ground state of a pre-quench Hermitian $XY$ Hamiltonian characterized by purely real initial parameters, specifically a transverse field $h_i$ and an anisotropy $\gamma_i$. For $t > 0$, the system evolves under the non-Hermitian Hamiltonian $\tilde{H}$. Because the evolution is generated by an effective non-Hermitian Hamiltonian, the norm of the state vector is not conserved. We therefore renormalize the state when evaluating physical observables.
Using the framework introduced in Sec.~\ref{sec:observable_difference}, in which right and left states are treated independently, the normalized time-evolved right and left states are defined as 
\begin{align}
    |\Psi^R(t)\rangle &= \frac{e^{-i \tilde{H} t} |\Psi_0\rangle}{\sqrt{\langle\Psi_0| e^{i \tilde{H}^\dagger t} e^{-i \tilde{H} t} |\Psi_0\rangle}}, \label{eq:psi_R} \\
    |\Psi^L(t)\rangle &= \frac{e^{-i \tilde{H}^\dagger t} |\Psi_0\rangle}{\sqrt{\langle\Psi_0| e^{i \tilde{H} t} e^{-i \tilde{H}^\dagger t} |\Psi_0\rangle}}. \label{eq:psi_L}
\end{align}

Because both the initial state and the post-quench Hamiltonian $\tilde{H}$ are quadratic in the fermionic operators, $|\Psi^R(t)\rangle$ and $|\Psi^L(t)\rangle$ remain Gaussian states at all times $t$. Consequently, the time-dependent state is fully parameterized by the evolution of the two-component spinors $\Phi_p(t) = (u_p(t), v_p(t))^T$ in each momentum sector $\phi_p$. Initialized by the pre-quench Hamiltonian as $\Phi_p(0)$, these spinors are independently propagated under the $2 \times 2$ post-quench block Hamiltonians $\tilde{H}_{\phi_p}$ and $\tilde{H}_{\phi_p}^\dagger$, 
\begin{align}
    \begin{pmatrix} u_p^R(t) \\ v_p^R(t) \end{pmatrix} &= \frac{1}{\mathcal{N}_p^R(t)} \exp(-i \tilde{H}_{\phi_p} t) \begin{pmatrix} u_p(0) \\ v_p(0) \end{pmatrix}, \\
    \begin{pmatrix} u_p^L(t) \\ v_p^L(t) \end{pmatrix} &= \frac{1}{\mathcal{N}_p^L(t)} \exp(-i \tilde{H}_{\phi_p}^\dagger t) \begin{pmatrix} u_p(0) \\ v_p(0) \end{pmatrix},
\end{align}
where the instantaneous normalization factors $\mathcal{N}_p^\nu(t)$ ensures $|u_p^\nu(t)|^2 + |v_p^\nu(t)|^2 = 1$. 
These time-dependent Bogoliubov amplitudes directly yield the dynamical momentum distributions $n^\nu(\phi_p, t) = |v_p^\nu(t)|^2$ and the anomalous pairing amplitudes $f^\nu(\phi_p, t) = (u_p^\nu(t))^* v_p^\nu(t)$. 

In what follows, we are interested in the entanglement entropy for a spatial subsystem $A$ of length $l$. To evaluate this, we must construct the real-space single-particle correlation matrices characterizing the subsystem \cite{vidal_lattore, Calabrese_2004}, 
%For spatial indices $m, n \in A$, the normal correlation matrix $\mathbf{C}^\nu(t)$ has elements $C_{mn}^\nu(t) = \langle c_m^\dagger c_n \rangle_\nu(t)$, and the anomalous correlation matrix $\mathbf{F}^\nu(t)$ has elements $F_{mn}^\nu(t) = \langle c_m c_n \rangle_\nu(t)$.
given by the normal correlation matrix $\mathbf{C}^\nu(t)$ with elements $C_{mn}^\nu(t) = \langle c_m^\dagger c_n \rangle_\nu(t)$ and the anomalous correlation matrix $\mathbf{F}^\nu(t)$ with elements $F_{mn}^\nu(t) = \langle c_m c_n \rangle_\nu(t)$, with spatial indices $m, n \in A$.

By substituting the inverse spatial Fourier transforms of the fermionic operators into these definitions, the matrix elements decouple into sums over the momentum-space contractions. Taking the thermodynamic limit, these discrete sums become continuous integrals over the Brillouin zone:
\begin{align}
    C_{mn}^\nu(t) &= \frac{1}{\pi} \int_0^\pi d\phi \cos[\phi(m-n)] n^\nu(\phi, t), \\
    F_{mn}^\nu(t) &= \frac{1}{\pi} \int_0^\pi d\phi \sin[\phi(m-n)] f^\nu(\phi, t).
\end{align}
Following the formalism introduced by Peschel \cite{Peschel_2003}, the reduced density matrix of the Gaussian subsystem is entirely encoded in the $2l \times 2l$ block covariance matrix
\begin{equation}
    \mathbf{\Gamma}^\nu_l(t) = 
    \begin{pmatrix}
    \mathbb{I}_{l \times l} - \mathbf{C}^\nu(t) & -\mathbf{F}^\nu(t) \\
    -(\mathbf{F}^\nu(t))^\dagger & (\mathbf{C}^\nu(t))^T 
    \end{pmatrix}.
\end{equation}
Diagonalizing $\mathbf{\Gamma}^\nu_l(t)$ yields $2l$ eigenvalues that, due to particle-hole symmetry, appear in pairs $\pm \eta_j^\nu(t)$ for $j = 1, \dots, l$. The von Neumann entanglement entropy for the left and right states at time $t$ is then analytically constructed from these single-particle pseudo-energies:
\begin{align}
\label{eq:entanglement_entropy}
    &S_l^\nu(t) = \\  &- \sum_{j=1}^{l} \left( \frac{1+\eta_j^\nu(t)}{2} \ln \frac{1+\eta_j^\nu(t)}{2} + \frac{1-\eta_j^\nu(t)}{2} \ln \frac{1-\eta_j^\nu(t)}{2} \right). \nonumber
\end{align}

By independently evaluating Eq.~\eqref{eq:entanglement_entropy} for the right ($\nu=R$) and left ($\nu=L$) state correlators, we can compute the entropy difference $S_l(t) \equiv S_l^R(t) - S_l^L(t)$ during the non-Hermitian quench dynamics. %We track the left and right time-evolved states, $\ket{\psi^L(t)}$ and $\ket{\psi^R(t)}$, and 
%In what follows, we compute the von Neumann entanglement entropy for spatial subsystems of varying lengths $l \in \{20, 40, 60, 80, 100\}$. 
Because the non-Hermiticity drives the left and right states toward distinct eigenmodes, the difference between their entanglement structures serves as a natural dynamical indicator for the exceptional point. We quantify this difference using the absolute instantaneous difference, $ |S^R_l(t) - S^L_l(t)|$, and its long-time average, $\overline{|S^R_l - S^L_l|} = \frac{1}{t_2 - t_1}\int_{t_1}^ {t_2}|S^R_l(t) - S^L_l(t)| dt$, where $t_1 = 50$ and $t_2 = 200$.
In what follows, we compute the von Neumann entanglement entropies for spatial subsystems of varying lengths $l \in \{20, 40, 60, 80, 100\}$.

\subsection{Entanglement difference in the $\mathcal{RT}$-Symmetric Regime}

We first consider the $\mathcal{RT}$-symmetric quench, where the post-quench non-Hermitian Hamiltonian is parameterized by a purely imaginary anisotropy ($\tilde{\gamma} = i\gamma^{\mathcal{RT}}$) and a real magnetic field ($\tilde{h} = h^{\mathcal{RT}}$). The numerical results for the time-averaged entanglement entropy difference $\overline{|S^R_l - S^L_l|}$ are presented in Fig.~\ref{fig:rt_sym_entro_difference} for varying anisotropy strengths $\gamma^{\mathcal{RT}} \in \{0.5, 0.8, 1\}$.

Across all values of $\gamma^{\mathcal{RT}}$, the time-averaged curves develop a pronounced peak that has a sharp flank at the exceptional point $h_{\mathrm{ep}}^{\mathcal{RT}} = \sqrt{1+(\gamma^{\mathcal{RT}})^2}$, indicated by the vertical dashed line. This alignment becomes increasingly sharp as the subsystem size $l$ increases. For smaller $l$, the peak remains visible but rounded, while for larger $l$ it emerges as a well-defined cusp. At the same time, the overall amplitude of the signal grows with subsystem size, indicating that the effect is macroscopic.

The time-resolved dynamics $|S^R_l(t) - S^L_l(t)|$ (Fig.~\ref{fig:rt_sym_entro_difference}, columns 2-4) reveal the underlying mechanisms driving the difference in behavior in the distinct parameter regimes:  
\begin{itemize}
    \item \textbf{Broken Phase ($h < h_{\mathrm{ep}}$):} When the system possesses a complex spectrum, the left and right states are projected toward different dominant eigenvectors corresponding to $H^{\mathcal{RT}}$ and ${H^{\mathcal{RT}}}^\dagger$. The entanglement difference exhibits an initial transient followed by long-lived oscillations with finite amplitude. These oscillations persist throughout the observed time window and directly translate into a finite contribution to the time-averaged quantity (see Fig.~\ref{fig:rt_sym_entro_difference}(a1), (b1), (c1) )
    
    \item \textbf{Exceptional Point ($h = h_{\mathrm{ep}}$):}   
    At the exceptional point, the dynamics change qualitatively. The entanglement difference exhibits sharp peaks at intermediate times that stand out clearly from the surrounding oscillations [e.g., Fig.~\ref{fig:rt_sym_entro_difference}(a2), (b2), (c2)]. These large transient contributions dominate the time average and produce the pronounced peak in $\overline{|S^R_l - S^L_l|}$.
    \item \textbf{Unbroken Phase ($h > h_{\mathrm{ep}}$):} In the parameter regime where the spectrum is entirely real, the oscillations are visibly damped. After the initial transient, the difference in entanglement steadily decreases and remains small at long times [e.g., Fig.~\ref{fig:rt_sym_entro_difference}(a3), (b3), (c3)].
\end{itemize}

\subsection{Entanglement difference as an order parameter in the $\mathcal{PT}$-Symmetric Regime}
\begin{figure*}
    \includegraphics[width=\linewidth]{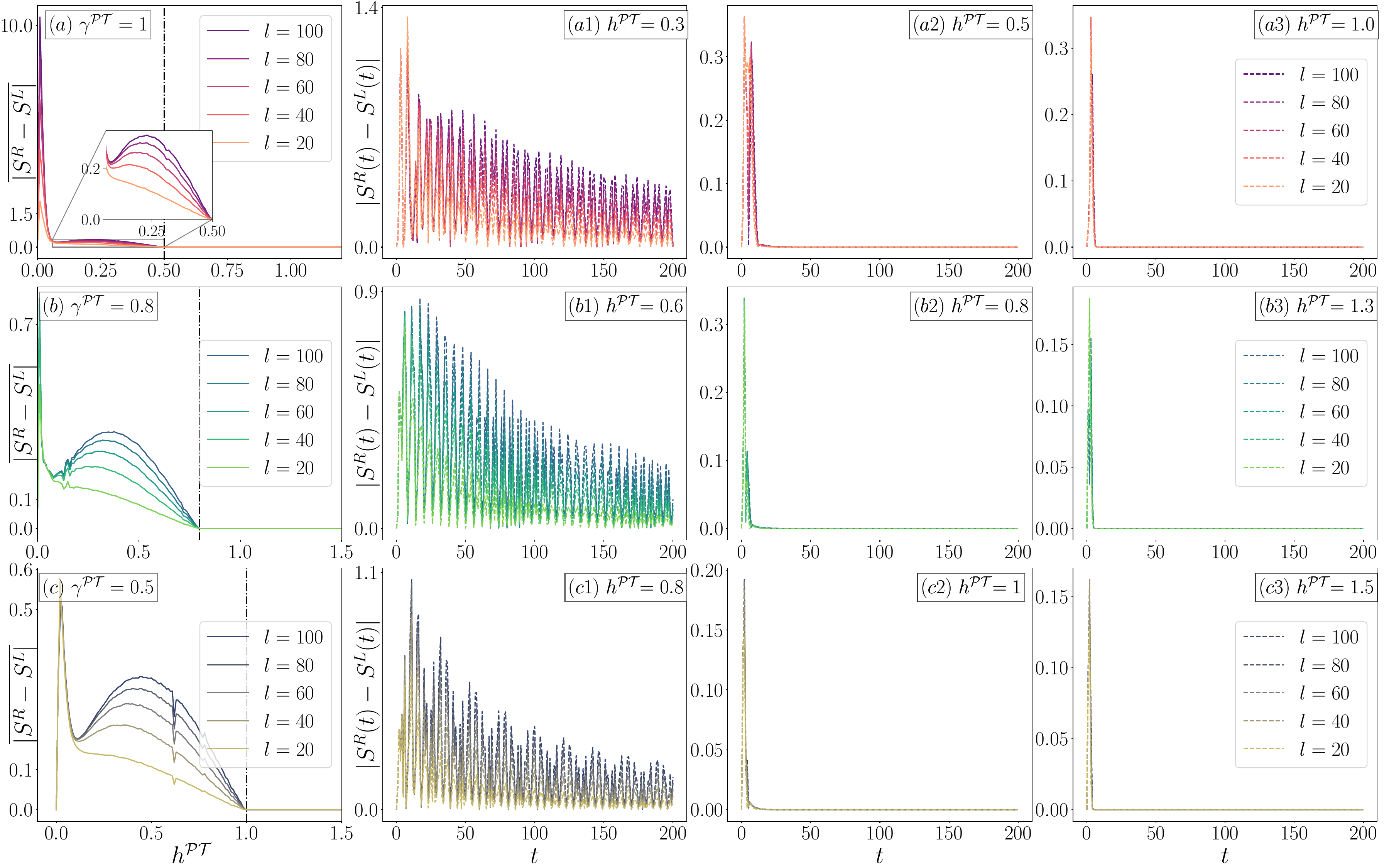}
    
    \caption{Dynamical entanglement entropy difference $|S_l^R - S_l^L|$ in the $\mathcal{PT}$-symmetric regime following a sudden quench. Panels in the left column [(a)–(c)] show the time-averaged quantity as a function of the imaginary transverse magnetic field $h^{PT}$ for subsystem sizes $l \in \{20,40,60,80,100\}$. The dash-dotted vertical lines indicate the corresponding exceptional points $h^{\mathcal{PT}}_{\mathrm{ep}} = \gamma^{\mathcal{PT}}$. The remaining panels [(a1)–(c3)] display the real-time evolution $|S_l^R(t) - S_l^L(t)|$ up to $t=200$ for representative values of $h^{\mathcal{PT}}$ indicated in each panel. For each row, the three panels correspond to field values below [(a1), (b1), (c1)], at [(a2), (b2), (c2)], and above [(a3), (b3), (c3)] the exceptional point. Rows correspond to anisotropy parameters $\gamma^{\mathcal{PT}} = 1$ [(a)], $0.8$ [(b)], and $0.5$ [(c)]. In the topologically non-trivial phase $(h^{\mathcal{PT}} < h^{\mathcal{PT}}_{\mathrm{ep}})$, the entanglement difference exhibits sustained oscillatory dynamics and remains finite after time averaging, indicating distinct dynamical evolution under the Hamiltonian and its adjoint. At the exceptional point $(h^{\mathcal{PT}} = h^{\mathcal{PT}}_{\mathrm{ep}})$, the dynamics display a sharp transient peak followed by rapid suppression of the left--right asymmetry. In the topologically trivial phase $(h^{\mathcal{PT}} > h^{\mathcal{PT}}_{\mathrm{ep}})$, the entanglement difference vanishes after the initial transient timescale, reflecting identical evolution of the right and left entanglement entropies. Colors indicate subsystem sizes $l$, from lighter to darker shades corresponding to smaller to larger $l$. All quantities are dimensionless.}
    \label{fig:pt_sym_entro_difference}
\end{figure*}
Next, we evaluate the quench dynamics into the $\mathcal{PT}$-symmetric regime, where the post-quench Hamiltonian is characterized by a real anisotropy ($\gamma^{\mathcal{PT}}$) and a purely imaginary magnetic field ($h^{\mathcal{PT}}$). The time-averaged entanglement entropy difference $\overline{|S_l^R - S_l^L|}$ and the corresponding real-time dynamics $|S_l^R(t) - S_l^L(t)|$ are presented in Fig.~\ref{fig:pt_sym_entro_difference} for varying anisotropy parameters $\gamma^{\mathcal{PT}} \in \{1, 0.8, 0.5\}$.

As shown in the leftmost column of Fig.~\ref{fig:pt_sym_entro_difference}, the time-averaged entanglement difference acts effectively as a dynamically generated order parameter for the topological phase transition. In the topologically non-trivial phase ($h^{\mathcal{PT}} < h_{\mathrm{ep}}^{\mathcal{PT}}$), the quantity $\overline{|S_l^R - S_l^L|}$ maintains a finite, size-dependent value. Upon crossing the exceptional point at $h_{\mathrm{ep}}^{\mathcal{PT}} = \gamma^{\mathcal{PT}}$, this quantity drops to zero and remains negligible throughout the topologically trivial phase ($h^{\mathcal{PT}} \ge h_{\mathrm{ep}}^{\mathcal{PT}}$).

The microscopic origin of this behavior is resolved by analyzing the time-resolved dynamics in the different topological regimes:
\begin{itemize}
    \item \textbf{In the topologically non-trivial phase ($h^{\mathcal{PT}} < h_{\mathrm{ep}}^{\mathcal{PT}}$):} As depicted in panels (a1), (b1), and (c1), the real-time difference $|S_l^R(t) - S_l^L(t)|$ exhibits slowly decaying, system-size-dependent oscillations. The time evolutions generated by the Hamiltonian and its adjoint drive the system toward different dynamical trajectories, resulting in a finite time-averaged entanglement difference.
    \item \textbf{At the exceptional point ($h^{\mathcal{PT}} = h_{\mathrm{ep}}^{\mathcal{PT}}$):} As shown in panels (a2), (b2), and (c2), the dynamics exhibit an initial, rapid transient peak. Following this brief initial response, the entanglement difference approaches zero.
    \item \textbf{In the topologically trivial phase ($h^{\mathcal{PT}} > h_{\mathrm{ep}}^{\mathcal{PT}}$):} As seen in panels (a3), (b3), and (c3), the real-time difference remains strictly zero after the initial transient timescale. In this regime, the purely real post-quench spectrum enforces an effectively unitary-like evolution, causing the right and left entanglement entropies to evolve identically and suppressing any dynamical asymmetry.
\end{itemize}

Comparing these results with the $\mathcal{RT}$-symmetric quench reveals a fundamental distinction in how the dynamical entanglement difference resolves the respective critical boundaries. In the $\mathcal{RT}$ regime, the maximum of the time-averaged asymmetry $\overline{|S_l^R - S_l^L|}$ drifts away from the exceptional point and into the bulk of the broken-symmetry phase for general anisotropies ($\gamma^{\mathcal{RT}} < 1$). In stark contrast, the $\mathcal{PT}$-symmetric quench exhibits no such parameter-dependent drift. Instead, the entropy difference acts as a strict, unambiguous indicator of the exceptional point across all considered values of $\gamma^{\mathcal{PT}}$. By approaching zero at $h_{\mathrm{ep}}^{\mathcal{PT}}$ and remaining negligibly small throughout the topologically trivial phase, the quantity $\overline{|S_l^R - S_l^L|}$ serves as a dynamically generated order-parameter-like indicator of the transition.

\section{Conclusion}
\label{sec:conclusion}

In this work, we have systematically investigated the exceptional behavior of the one-dimensional non-Hermitian $XY$ spin chain across both its $\mathcal{RT}$- and $\mathcal{PT}$-symmetric regimes. By combining a Hamiltonian-based trace-norm quantifier with local observable-based diagnostics, we established a consistent set of experimentally accessible signatures for detecting exceptional points and resolving the associated exceptional boundaries. The Hamiltonian quantifier $\mathcal{D}$ exhibits distinct non-analytic features that depend on the underlying symmetry class, with a cusp in its first derivative signaling the $\mathcal{RT}$-symmetry-breaking transition and a divergence in the second derivative identifying the $\mathcal{PT}$-symmetric topological transition.

At the level of observables, evaluating local spin correlators independently in the right and left ground-state manifolds revealed that differences between the two provide sensitive probes of these exceptional boundaries. Importantly, this sensitivity is highly selective: while certain local observables display clear signatures of the transition, static bipartite entanglement measures remain identical in the two manifolds across all regimes, indicating that they do not capture the underlying non-Hermitian structure.

This limitation of static probes is overcome by considering non-equilibrium dynamics. Following a sudden quench into the non-Hermitian regime, the evolutions generated by the Hamiltonian and its adjoint give rise to distinct dynamical trajectories, which in turn produce a finite entanglement entropy difference. In the $\mathcal{RT}$-symmetric regime, this quantity exhibits a pronounced maximum at the exceptional point. In contrast, in the $\mathcal{PT}$-symmetric regime, the time-averaged entanglement entropy difference $\overline{|S^R_l - S^L_l|}$ behaves as an order-parameter-like quantity that drops sharply to zero at the exceptional point, providing a direct dynamical signature of the transition.

Taken together, these results establish a coherent framework in which Hamiltonian-level quantifiers, observable-based diagnostics, and dynamical probes provide complementary insights into non-Hermitian many-body systems. In particular, they show that while the imprint of non-Hermiticity on static observables can be subtle and selective, non-equilibrium dynamics offer a physically transparent route to detecting exceptional points. This interplay between static and dynamical signatures opens up a pathway for probing non-Hermitian physics in quantum simulators and for exploiting such systems in controlled state preparation and quantum information applications.

\acknowledgments

This project has been funded by the Caritro Foundation. This work was supported by the Provincia Autonoma di Trento, and Q@TN, the joint lab between the University of Trento, FBK—Fondazione Bruno Kessler, INFN—National Institute for Nuclear Physics, and CNR—National Research Council, Italy. S.S.R. acknowledges financial support from the Faculty Research Scheme, IIT (ISM) Dhanbad, India, under Project No. FRS/2024/PHYSICS/MISC0110, and from the Anusandhan National Research Foundation (ANRF), Government of India, under Grant Nos.\ ANRF/ARG/2025/004617/PS and ANRF/ECRG/2025/002793/PMS.

\appendix
\section{$H^{\mathcal{RT}}$ as an effective Hamiltonian via Reservoir engineering }
\label{sec:app_rt_derivation}

In this appendix, we present a detailed derivation of the $\mathcal{RT}$-symmetric non-Hermitian $XY$ model with purely imaginary anisotropy, starting from a Hermitian spin system coupled to engineered dissipative environments. This construction provides a transparent microscopic origin of the effective non-Hermitian Hamiltonian employed in the main text and clarifies how imaginary anisotropic couplings naturally emerge from open-system dynamics.

As our starting point, we consider a one-dimensional chain of spin-$\frac{1}{2}$ particles with nearest-neighbor interactions, described by the Hermitian Hamiltonian
\begin{equation}
H_S = J \sum_j \left( \sigma_j^x \sigma_{j+1}^x + \sigma_j^y \sigma_{j+1}^y \right) + \frac{h^{\mathcal{RT}}}{2} \sum_j \sigma_j^z,
\end{equation}
where $\sigma_j^\alpha$ $(\alpha = x,y,z)$ denote the Pauli matrices, $J$ is the exchange interaction strength, $h$ is the transverse magnetic field, and subscript $S$ indicates the system of interest. In the following, we set $J=1$ for simplicity.

We assume that the system is weakly coupled to an environment, such that its dynamics is governed by a Markovian master equation of Lindblad form,
\begin{equation}
\frac{d\rho}{dt} = -i[H_S,\rho] + \sum_j \left( L_j \rho L_j^\dagger - \frac{1}{2} \{L_j^\dagger L_j, \rho\} \right),
\end{equation}
which arises under the standard Born–Markov approximation for open quantum systems \cite{Breuer_Petruccione_book}. Here, $L_j$ denote jump operators that encode the coupling to the environment.

To generate non-Hermitian interactions, we consider dissipative processes that act non-locally on neighboring spins. In particular, we introduce jump operators of the form
\begin{equation}
L_j = p\, \sigma_j^- + q\, \sigma_j^+ + r\, \sigma_{j+1}^- + s\, \sigma_{j+1}^+,
\end{equation}
where $\sigma_j^{\pm} = (\sigma_j^x \pm i \sigma_j^y)/2$, and the coefficients $p,q,r,s \in \mathbb{C}$ characterize the strength and structure of system–reservoir couplings \cite{zoller_reservoir_engineering_1996}, which can be used to engineer or modify dissipative processes to obtain desired quantum behavior \cite{giorgi_dissipative_resource_quantum_reservoir_engineering} such as nonreciprocal photon transmission \cite{clerk_reservoir_engineering_non_reciprocal_2015}, persistent currents \cite{fazio_persistent_current_reservoir_engineering_2018}, 
unequal time anti-commutators~\cite{kevin_hauke_prxq_2022}, imaginary hoppings to non-invasively measure currents \cite{geier2021noninvasivemeasurementcurrentsanalog}. 

We now specialize to a class of jump operators that induce correlated gain and loss processes between neighboring sites. Specifically, we set $p = s = 0$ and consider $
L_j = q\, \sigma_j^+ + r\, \sigma_{j+1}^-,
$
with complex coefficients $q = -\sqrt{\gamma^{\mathcal{RT}}/2}$ and $r = \sqrt{\gamma^{\mathcal{RT}}/2}$. This choice generates asymmetric pairing processes and is sufficient to produce the desired non-Hermitian anisotropy.

We resort to a quantum trajectory framework, where the evolution conditioned on the absence of quantum jumps is governed by an effective non-Hermitian Hamiltonian, known as the no-click limit \cite{quantum_jump_plenio,Daley_review_quantum_jumps,turkeshi_no_click_limit_2021}
\begin{equation}
H_{\mathrm{eff}} = H_S - \frac{i}{2} \sum_j L_j^\dagger L_j.
\end{equation}
The non-Hermitian contribution originates from the term $-\frac{i}{2} L_j^\dagger L_j$, which encodes the postselection induced loss of probability. Evaluating $L_j^\dagger L_j$ explicitly, we obtain
\begin{equation}
L_j^\dagger L_j 
= \frac{\gamma^{\mathcal{RT}}}{4}( \sigma_j^- \sigma_j^+ 
+ \sigma_{j+1}^+ \sigma_{j+1}^- -  \sigma_j^- \sigma_{j+1}^- - \sigma_{j+1}^+ \sigma_j^+) .
\end{equation}
We express the spin ladder operators in terms of Pauli matrices,
$
\sigma_j^\pm = \frac{1}{2}(\sigma_j^x \pm i \sigma_j^y).
$ Here, the on-site terms reduce to
\begin{equation}
\sigma_j^- \sigma_j^+ = \frac{1}{2}(1 - \sigma_j^z), 
\qquad
\sigma_{j+1}^+ \sigma_{j+1}^- = \frac{1}{2}(1 + \sigma_{j+1}^z),
\end{equation}
which only contribute local shifts (a constant and a renormalization of the magnetic field). These terms do not generate any anisotropic coupling between different sites.

We now focus on the cross terms. Using the definitions above, we compute
\begin{align}
\sigma_j^- \sigma_{j+1}^- 
%&= \frac{1}{4}(\sigma_j^x - i\sigma_j^y)(\sigma_{j+1}^x - i\sigma_{j+1}^y) \nonumber \\
&= \frac{1}{4} \left[
\sigma_j^x \sigma_{j+1}^x 
- i \sigma_j^x \sigma_{j+1}^y
- i \sigma_j^y \sigma_{j+1}^x
- \sigma_j^y \sigma_{j+1}^y
\right],
\end{align}
and similarly,
\begin{align}
\sigma_{j+1}^+ \sigma_j^+ 
%&= \frac{1}{4}(\sigma_{j+1}^x + i\sigma_{j+1}^y)(\sigma_j^x + i\sigma_j^y) \nonumber \\
&= \frac{1}{4} \left[
\sigma_j^x \sigma_{j+1}^x 
+ i \sigma_j^x \sigma_{j+1}^y
+ i \sigma_j^y \sigma_{j+1}^x
- \sigma_j^y \sigma_{j+1}^y
\right],
\end{align}
where in the second step we used the fact that operators on different sites commute. Adding these two expressions, we obtain a complete cancellation of the mixed terms,
\begin{align}
\sigma_j^- \sigma_{j+1}^- + \sigma_{j+1}^+ \sigma_j^+ 
%&= \frac{1}{4} \Big[2\,\sigma_j^x \sigma_{j+1}^x - 2\,\sigma_j^y \sigma_{j+1}^y\Big] \nonumber \\
&= \frac{1}{2}\left(
\sigma_j^x \sigma_{j+1}^x 
- \sigma_j^y \sigma_{j+1}^y
\right).
\end{align}

We thus see explicitly that the dissipative processes generate an anisotropic interaction in the $XY$ plane, with opposite contributions to the $xx$ and $yy$ couplings. When combined with the isotropic $XX$ interaction in $H_S$, this yields the effective Hamiltonian \cite{Song_RT_symm,recog_keshav_tanoy_aditi}
\begin{align}
H_{\mathrm{eff}} \equiv H^{\mathcal{RT}} &=  \sum_j \left[
\frac{1 + i\gamma^\mathcal{RT}}{4} \, \sigma_j^x \sigma_{j+1}^x
+ \frac{1 - i\gamma^\mathcal{RT}}{4} \, \sigma_j^y \sigma_{j+1}^y \right. \nonumber \\
&+ \left. \frac{h^{\mathcal{RT}}}{2}\sigma_j^z
\right],
\end{align}
which we called as $H^\mathcal{RT}$ in the main text.

This construction demonstrates that the $\mathcal{RT}$-symmetric $XY$ model with imaginary anisotropy naturally emerges from a Hermitian spin chain subject to reservoir engineered dissipative processes, thereby providing a physically grounded starting point for its analysis.
\section{$H^{\mathcal{PT}}$ as an effective Hamiltonian for the decay of a three level atom }
\label{app:pt_sym_derivation}

In this appendix, we derive the $\mathcal{PT}$-symmetric non-Hermitian $XY$ model from a system of decaying three-level atoms. In this framework, the non-Hermitian Hamiltonian arises by conditioning the dynamics on no decay events, resulting in an effective imaginary magnetic field \cite{heralded_pt_, PT_sym_imaginary_magnetic_field}. 

We consider a one-dimensional array of atoms, each of which possesses three relevant internal states: a ground state $\ket{\downarrow}$, an excited state $\ket{\uparrow}$, and an auxiliary state $|a\rangle$. The two states $\ket{\uparrow}$ and $\ket{\downarrow}$ define an effective spin-$\frac{1}{2}$ degree of freedom, while the state $|a\rangle$ serves as a decay channel. Physically, the system is characterized by spontaneous decay processes of the form
\begin{equation}
\ket{\uparrow} \;\longrightarrow\; |a\rangle,
\end{equation}
which occur at a rate $\gamma$.

In addition to this dissipative process, the spins interact coherently via an anisotropic $XY$ Hamiltonian,
\begin{equation}
H = \sum_j \left( \frac{1+\gamma^{\mathcal{PT}}}{4} \, \sigma_j^x \sigma_{j+1}^x + \frac{1-\gamma^{\mathcal{PT}}}{4} \sigma_j^y \sigma_{j+1}^y \right),
\end{equation}
where $(1\pm\gamma^{\mathcal{PT}})/{4}$ are real coupling strengths. This coherent part tends to generate spin excitations and correlations, while the decay process continuously removes population from the $\ket{\uparrow}$ state.

The crucial ingredient is that the environment is continuously monitoring whether a decay event occurs. In each experimental realization, one post-selects trajectories in which no decay is observed. As a result, the system evolves under a conditional (no-jump) dynamics which is no longer unitary \cite{quantum_jump_plenio,Daley_review_quantum_jumps,turkeshi_no_click_limit_2021}. Instead, the evolution is governed by an effective non-Hermitian Hamiltonian 
\begin{equation}
H_{\mathrm{eff}} = H - \frac{i}{2} \sum_j L_j^\dagger L_j,
\end{equation}
where $L_j$ are the jump operators associated with the decay processes.

For the spontaneous decay $\ket{\uparrow} \to |a\rangle$, the corresponding jump operator is given by
\begin{equation}
L_j = \sqrt{2h^{\mathcal{PT}}}\sigma_j^-.
\end{equation}
The operator $L_j^\dagger L_j$ determines the rate at which probability is removed from the system due to decay. Evaluating this operator, we obtain
\begin{equation}
L_j^\dagger L_j = 2h^{\mathcal{PT}} \sigma_j^+ \sigma_j^-.
\end{equation}
Using the identity
\begin{equation}
\sigma_j^+ \sigma_j^- = \frac{1}{2}(1 + \sigma_j^z),
\end{equation}
the non-Hermitian contribution can be written as
\begin{equation}
-\frac{i}{2} L_j^\dagger L_j = -\frac{ih^{\mathcal{PT}}}{2}(1 + \sigma_j^z).
\end{equation}
This term has a simple physical interpretation. Since $\sigma_j^z = +1$ for the state $|\uparrow\rangle$ and $\sigma_j^z = -1$ for $|\downarrow\rangle$, the imaginary term selectively suppresses configurations with spin-up excitations. In other words, even in the absence of an actual decay event, the continuous monitoring induces a bias in the dynamics that penalizes the population of the decaying state $|\uparrow\rangle$. This reflects the fact that trajectories containing more $|\uparrow\rangle$ components are more likely to decay and, therefore, are less likely to survive the post-selection.

Substituting into the effective Hamiltonian and neglecting a constant shift that does not affect the dynamics or spectral properties, we obtain
\begin{align}
H_{\mathrm{eff}} \equiv H^{\mathcal{PT}} &=  \sum_j \left( \frac{1+\gamma^{\mathcal{PT}}}{4} \, \sigma_j^x \sigma_{j+1}^x + \frac{1-\gamma^{\mathcal{PT}}}{4} \, \sigma_j^y \sigma_{j+1}^y \right) \nonumber \\
&- \frac{ih^{\mathcal{PT}}}{2} \sum_j \sigma_j^z
\end{align}
corresponds to a non-Hermitian $XY$ model with real anisotropy and a purely imaginary transverse magnetic field.

In this way, the $\mathcal{PT}$-symmetric $XY$ model arises naturally from a system of interacting spins subject to decay and continuous observation, with the imaginary magnetic field encoding the measurement-induced suppression of unstable states.

\bibliography{references}

@article{Peschel_2003,
doi = {10.1088/0305-4470/36/14/101},
url = {https://doi.org/10.1088/0305-4470/36/14/101},
year = {2003},
month = {mar},
publisher = {},
volume = {36},
number = {14},
pages = {L205},
author = {Ingo Peschel},
title = {Calculation of reduced density matrices from correlation functions},
journal = {Journal of Physics A: Mathematical and General}
}

@article{pt_symmetry_experiment_floquet_joglekar,
	author = {Li, Jiaming and Harter, Andrew K. and Liu, Ji and de Melo, Leonardo and Joglekar, Yogesh N. and Luo, Le},
	title = {{Observation of parity-time symmetry breaking transitions in a dissipative Floquet system of ultracold atoms}},
	journal = {Nat Commun},
	volume = {10},
	number = {855},
	pages = {855},
	year = {2019},
	month = feb,
	issn = {2041-1723},
	publisher = {Nature Publishing Group},
	doi = {10.1038/s41467-019-08596-1}
}

@article{pt_symmetry_dynamics_photons,
  title={Experimental investigation of the no-signalling principle in parity--time symmetric theory using an open quantum system},
  author={Tang, Jian-Shun and Wang, Yi-Tao and Yu, Shang and He, De-Yong and Xu, Jin-Shi and Liu, Bi-Heng and Chen, Geng and Sun, Yong-Nan and Sun, Kai and Han, Yong-Jian and others},
  journal={Nature Photonics},
  volume={10},
  number={10},
  pages={642--646},
  year={2016},
  publisher={Nature Publishing Group}
}

@article{pt_dynamics_experiment_photonics,
author = {Wei-Chao Gao and Chao Zheng and Lu Liu and Tie-Jun Wang and Chuan Wang},
journal = {Opt. Express},
number = {1},
pages = {517--526},
publisher = {Optica Publishing Group},
title = {Experimental simulation of the parity-time symmetric dynamics using photonic qubits},
volume = {29},
month = {Jan},
year = {2021},
url = {https://opg.optica.org/oe/abstract.cfm?URI=oe-29-1-517},
doi = {10.1364/OE.405815},
}

@article{pt_symmetry_dynamics_photonics_topological_edge,
	author = {Xiao, L. and Zhan, X. and Bian, Z. H. and Wang, K. K. and Zhang, X. and Wang, X. P. and Li, J. and Mochizuki, K. and Kim, D. and Kawakami, N. and Yi, W. and Obuse, H. and Sanders, B. C. and Xue, P.},
	title = {{Observation of topological edge states in parity-time-symmetric quantum walks}},
	journal = {Nat Phys},
	volume = {13},
	pages = {1117--1123},
	year = {2017},
	month = nov,
	issn = {1745-2481},
	publisher = {Nature Publishing Group},
	doi = {10.1038/nphys4204}
}

@article{pt_symmetry_dynamics_nitrogen_vacancy,
	author = {Wu, Yang and Liu, Wenquan and Geng, Jianpei and Song, Xingrui and Ye, Xiangyu and Duan, Chang-Kui and Rong, Xing and Du, Jiangfeng},
	title = {{Observation of parity-time symmetry breaking in a single-spin system}},
	journal = {Science},
	volume = {364},
	number = {6443},
	pages = {878--880},
	year = {2019},
	month = may,
	issn = {0036-8075},
	publisher = {American Association for the Advancement of Science},
	doi = {10.1126/science.aaw8205}
}

@article{Song_RT_symm,
  title = {Non-Hermitian anisotropic $XY$ model with intrinsic rotation-time-reversal symmetry},
  author = {Zhang, X. Z. and Song, Z.},
  journal = {Phys. Rev. A},
  volume = {87},
  issue = {1},
  pages = {012114},
  numpages = {7},
  year = {2013},
  month = {Jan},
  publisher = {American Physical Society},
  doi = {10.1103/PhysRevA.87.012114},
  url = {https://link.aps.org/doi/10.1103/PhysRevA.87.012114}
}

@article{Calabrese_2004,
doi = {10.1088/1742-5468/2004/06/P06002},
url = {https://dx.doi.org/10.1088/1742-5468/2004/06/P06002},
year = {2004},
month = {jun},
publisher = {},
volume = {2004},
number = {06},
pages = {P06002},
author = {Pasquale Calabrese and John Cardy},
title = {Entanglement entropy and quantum field theory},
journal = {Journal of Statistical Mechanics: Theory and Experiment},
}

@incollection{jordan-wigner,
  title={{\"u}ber das paulische {\"a}quivalenzverbot},
  author={Jordan, Pascual and Wigner, Eugene Paul},
  booktitle={The Collected Works of Eugene Paul Wigner},
  pages={109--129},
  year={1993},
  publisher={Springer}
}

@article{bm1,
  title = {Statistical Mechanics of the $\mathrm{XY}$ Model. I},
  author = {Barouch, Eytan and McCoy, Barry M. and Dresden, Max},
  journal = {Phys. Rev. A},
  volume = {2},
  issue = {3},
  pages = {1075--1092},
  numpages = {0},
  year = {1970},
  month = {Sep},
  publisher = {American Physical Society},
  doi = {10.1103/PhysRevA.2.1075},
  url = {https://link.aps.org/doi/10.1103/PhysRevA.2.1075}
}

@Article{ising_scipost_santoro,
	title={{The quantum Ising chain for beginners}},
	author={Glen Bigan Mbeng and Angelo Russomanno and Giuseppe E. Santoro},
	journal={SciPost Phys. Lect. Notes},
	pages={82},
	year={2024},
	publisher={SciPost},
	doi={10.21468/SciPostPhysLectNotes.82},
	url={https://scipost.org/10.21468/SciPostPhysLectNotes.82},
}

@article{PT_sym_imaginary_magnetic_field,
doi = {10.1088/1361-648X/ac06ef},
url = {https://doi.org/10.1088/1361-648X/ac06ef},
year = {2021},
month = {jun},
publisher = {IOP Publishing},
volume = {33},
number = {34},
pages = {345601},
author = {Pi, Jinghui and Lü, Rong},
title = {Phase diagram and quantum criticality of a non-Hermitian XY model with a complex transverse field},
journal = {Journal of Physics: Condensed Matter}
}

@article{vidal_lattore,
  title = {Entanglement in Quantum Critical Phenomena},
  author = {Vidal, G. and Latorre, J. I. and Rico, E. and Kitaev, A.},
  journal = {Phys. Rev. Lett.},
  volume = {90},
  issue = {22},
  pages = {227902},
  numpages = {4},
  year = {2003},
  month = {Jun},
  publisher = {American Physical Society},
  doi = {10.1103/PhysRevLett.90.227902},
  url = {https://link.aps.org/doi/10.1103/PhysRevLett.90.227902}
}

@book{Breuer_Petruccione_book,
  doi = {10.1093/acprof:oso/9780199213900.001.0001},
  url = {https://doi.org/10.1093/acprof:oso/9780199213900.001.0001},
  year = {2007},
  month = jan,
  publisher = {Oxford University Press},
  author = {Heinz-Peter Breuer and Francesco Petruccione},
  title = {The Theory of Open Quantum Systems}
}

@article{recog_keshav_tanoy_aditi,
  title = {Recognizing critical lines via entanglement in non-Hermitian systems},
  author = {Das Agarwal, Keshav and Konar, Tanoy Kanti and Lakkaraju, Leela Ganesh Chandra and Sen(De), Aditi},
  journal = {Phys. Rev. A},
  volume = {113},
  issue = {2},
  pages = {022201},
  numpages = {14},
  year = {2026},
  month = {Feb},
  publisher = {American Physical Society},
  doi = {10.1103/b5sd-fn57},
  url = {https://link.aps.org/doi/10.1103/b5sd-fn57}
}

@misc{agarwal2025criticalquantummetrologyusing,
      title={Critical quantum metrology using non-Hermitian spin model with RT-symmetry}, 
      author={Keshav Das Agarwal and Tanoy Kanti Konar and Leela Ganesh Chandra Lakkaraju and Aditi Sen De},
      year={2025},
      eprint={2503.24331},
      archivePrefix={arXiv},
      primaryClass={quant-ph},
      url={https://arxiv.org/abs/2503.24331}, 
}

@article{non_herm_battery_2024,
  title = {Quantum battery with non-Hermitian charging},
  author = {Konar, Tanoy Kanti and Lakkaraju, Leela Ganesh Chandra and Sen (De), Aditi},
  journal = {Phys. Rev. A},
  volume = {109},
  issue = {4},
  pages = {042207},
  numpages = {12},
  year = {2024},
  month = {Apr},
  publisher = {American Physical Society},
  doi = {10.1103/PhysRevA.109.042207},
  url = {https://link.aps.org/doi/10.1103/PhysRevA.109.042207}
}

@misc{agarwal2026quantumenhancedsensinginterplaylongrange,
      title={Quantum-enhanced sensing from the interplay of long-range interactions and non-Hermiticity}, 
      author={Keshav Das Agarwal and Tanoy Kanti Konar and Leela Ganesh Chandra Lakkaraju and Aditi Sen De},
      year={2026},
      eprint={2605.01912},
      archivePrefix={arXiv},
      primaryClass={quant-ph},
      url={https://arxiv.org/abs/2605.01912}, 
}

@article{non_herm_extension_batchelor_2025,
  title = {Characterizing phase transitions and criticality in non-Hermitian extensions of the XY model},
  author = {Liu, D. C. and Batchelor, Murray T.},
  journal = {Phys. Rev. B},
  volume = {112},
  issue = {1},
  pages = {014422},
  numpages = {11},
  year = {2025},
  month = {Jul},
  publisher = {American Physical Society},
  doi = {10.1103/b55l-7tbc},
  url = {https://link.aps.org/doi/10.1103/b55l-7tbc}
}

@article{non_hermitian_winding_number_chaohong_2020,
  title = {Dynamic winding number for exploring band topology},
  author = {Zhu, Bo and Ke, Yongguan and Zhong, Honghua and Lee, Chaohong},
  journal = {Phys. Rev. Res.},
  volume = {2},
  issue = {2},
  pages = {023043},
  numpages = {12},
  year = {2020},
  month = {Apr},
  publisher = {American Physical Society},
  doi = {10.1103/PhysRevResearch.2.023043},
  url = {https://link.aps.org/doi/10.1103/PhysRevResearch.2.023043}
}

@article{bender_prl_1998,
  title = {Real Spectra in Non-Hermitian Hamiltonians Having $\mathcal{PT}$ Symmetry},
  author = {Bender, Carl M. and Boettcher, Stefan},
  journal = {Phys. Rev. Lett.},
  volume = {80},
  issue = {24},
  pages = {5243--5246},
  numpages = {0},
  year = {1998},
  month = {Jun},
  publisher = {American Physical Society},
  doi = {10.1103/PhysRevLett.80.5243},
  url = {https://link.aps.org/doi/10.1103/PhysRevLett.80.5243}
}

@article{pt_symmetry_dynamics_superconducting,
  title = {Experimental demonstration of a digital quantum simulation of a general $\mathcal{PT}$-symmetric system},
  author = {Wen, Jingwei and Zheng, Chao and Kong, Xiangyu and Wei, Shijie and Xin, Tao and Long, Guilu},
  journal = {Phys. Rev. A},
  volume = {99},
  issue = {6},
  pages = {062122},
  numpages = {8},
  year = {2019},
  month = {Jun},
  publisher = {American Physical Society},
  doi = {10.1103/PhysRevA.99.062122},
  url = {https://link.aps.org/doi/10.1103/PhysRevA.99.062122}
}

@book{exceptional_point_textbook,
  title={Perturbation Theory for Linear Operators},
  author={Kato, Tosio},
  year={1966},
  publisher={Springer},
  address={New York}
}

@article{lin2011unidirectional,
  title={Unidirectional Invisibility Induced by $\mathcal{PT}$-Symmetric Periodic Structures},
  author={Lin, Zin and Ramezani, Hamid and Eichelkraut, Toni and Kottos, Tsampikos and Cao, Hui and Christodoulides, Demetrios N.},
  journal={Physical Review Letters},
  volume={106},
  number={21},
  pages={213901},
  year={2011},
  publisher={American Physical Society},
  doi={10.1103/PhysRevLett.106.213901},
  url={https://doi.org/10.1103/PhysRevLett.106.213901}
}

@article{feng2013experimental,
  title={Experimental demonstration of a unidirectional reflectionless parity-time metamaterial at optical frequencies},
  author={Feng, Liang and Xu, Yuan-Lin and Fegadolli, William S. and Lu, Ming-Hui and Oliveira, Jos{\'e} E. B. and Almeida, Vilson R. and Chen, Yan-Feng and Scherer, Axel},
  journal={Nature Materials},
  volume={12},
  number={2},
  pages={108--113},
  year={2013},
  publisher={Nature Publishing Group},
  doi={10.1038/nmat3495},
  url={https://doi.org/10.1038/nmat3495}
}

@article{longhi2010pt,
  title={$\mathcal{PT}$-symmetric laser absorber},
  author={Longhi, Stefano},
  journal={Physical Review A},
  volume={82},
  number={3},
  pages={031801},
  year={2010},
  publisher={American Physical Society},
  doi={10.1103/PhysRevA.82.031801},
  url={https://doi.org/10.1103/PhysRevA.82.031801}
}

@article{chong2011pt,
  title={$\mathcal{PT}$-Symmetry Breaking and Laser-Absorber Modes in Optical Scattering Systems},
  author={Chong, Y. D. and Ge, Li and Stone, A. Douglas},
  journal={Physical Review Letters},
  volume={106},
  number={9},
  pages={093902},
  year={2011},
  publisher={American Physical Society},
  doi={10.1103/PhysRevLett.106.093902},
  url={https://doi.org/10.1103/PhysRevLett.106.093902}
}

@article{harari2018topological,
  title={Topological insulator laser: Theory},
  author={Harari, Gal and Bandres, Miguel A. and Lumer, Yaakov and Rechtsman, Mikael C. and Chong, Y. D. and Khajavikhan, Mercedeh and Christodoulides, Demetrios N. and Segev, Mordechai},
  journal={Science},
  volume={359},
  number={6381},
  pages={eaar4003},
  year={2018},
  publisher={American Association for the Advancement of Science},
  doi={10.1126/science.aar4003},
  url={https://doi.org/10.1126/science.aar4003}
}

@article{bandres2018topological,
  title={Topological insulator laser: Experiments},
  author={Bandres, Miguel A. and Wittek, Steffen and Harari, Gal and Parto, Midya and Ren, Jinhan and Segev, Mordechai and Christodoulides, Demetrios N. and Khajavikhan, Mercedeh},
  journal={Science},
  volume={359},
  number={6381},
  pages={eaar4005},
  year={2018},
  publisher={American Association for the Advancement of Science},
  doi={10.1126/science.aar4005},
  url={https://doi.org/10.1126/science.aar4005}
}

@article{hodaei2017enhanced,
  title={Enhanced sensitivity at higher-order exceptional points},
  author={Hodaei, Hossein and Hassan, Absar U. and Wittek, Steffen and Garcia-Gracia, Hipolito and El-Ganainy, Ramy and Christodoulides, Demetrios N. and Khajavikhan, Mercedeh},
  journal={Nature},
  volume={548},
  number={7666},
  pages={187--191},
  year={2017},
  publisher={Nature Publishing Group},
  doi={10.1038/nature23280},
  url={https://doi.org/10.1038/nature23280}
}

@article{chen2017exceptional,
  title={Exceptional points enhance sensing in an optical microcavity},
  author={Chen, Weijian and Kaya {\"O}zdemir, {\c{S}}ahin and Zhao, Guangming and Wiersig, Jan and Yang, Lan},
  journal={Nature},
  volume={548},
  number={7666},
  pages={192--196},
  year={2017},
  publisher={Nature Publishing Group},
  doi={10.1038/nature23281},
  url={https://doi.org/10.1038/nature23281}
}

@article{moiseyev2011non,
  title={Non-Hermitian Quantum Mechanics},
  author={Moiseyev, Nimrod},
  journal={Physics Reports},
  volume={302},
  number={5-6},
  pages={212--293},
  year={1998},
  publisher={Elsevier},
  doi={10.1016/S0370-1573(98)00002-7},
  url={https://doi.org/10.1016/S0370-1573(98)00002-7}
}

@article{dicke1954coherence,
  title={Coherence in Spontaneous Radiation Processes},
  author={Dicke, R. H.},
  journal={Physical Review},
  volume={93},
  number={1},
  pages={99--110},
  year={1954},
  publisher={American Physical Society},
  doi={10.1103/PhysRev.93.99},
  url={https://doi.org/10.1103/PhysRev.93.99}
}

@article{facchi2008quantum,
  title={Quantum Zeno dynamics: mathematical and physical aspects},
  author={Facchi, P and Pascazio, S},
  journal={Journal of Physics A: Mathematical and Theoretical},
  volume={41},
  number={49},
  pages={493001},
  year={2008},
  publisher={IOP Publishing},
  doi={10.1088/1751-8113/41/49/493001},
  url={https://doi.org/10.1088/1751-8113/41/49/493001}
}

@article{ashida2017parity,
  title={Parity-time-symmetric quantum critical phenomena},
  author={Ashida, Yuto and Furukawa, Shunsuke and Ueda, Masahito},
  journal={Nature Communications},
  volume={8},
  number={1},
  pages={15791},
  year={2017},
  publisher={Nature Publishing Group},
  doi={10.1038/ncomms15791},
  url={https://doi.org/10.1038/ncomms15791}
}

@article{Heiss_physics_of_exceptional_points,
	author = {Heiss, W. D.},
	title = {{The physics of exceptional points}},
	journal = {J Phys A: Math Theor},
	volume = {45},
	number = {44},
	pages = {444016},
	year = {2012},
	month = oct,
	issn = {1751-8121},
	publisher = {IOP Publishing},
	doi = {10.1088/1751-8113/45/44/444016}
}

@article{bender_ropp_2007,
doi = {10.1088/0034-4885/70/6/R03},
url = {https://dx.doi.org/10.1088/0034-4885/70/6/R03},
year = {2007},
month = {may},
publisher = {},
volume = {70},
number = {6},
pages = {947},
author = {Carl M Bender},
title = {Making sense of non-Hermitian Hamiltonians},
journal = {Reports on Progress in Physics}
}

@article{brody_iop_2014,
doi = {10.1088/1751-8113/47/3/035305},
url = {https://dx.doi.org/10.1088/1751-8113/47/3/035305},
year = {2013},
month = {dec},
publisher = {IOP Publishing},
volume = {47},
number = {3},
pages = {035305},
author = {Dorje C Brody},
title = {Biorthogonal quantum mechanics},
journal = {Journal of Physics A: Mathematical and Theoretical}
}

@article{ground_state_zhang_2022,
title = {Parameter estimation with the steady states of non-Hermitian spin chains},
journal = {Physica A: Statistical Mechanics and its Applications},
volume = {599},
pages = {127460},
year = {2022},
issn = {0378-4371},
doi = {https://doi.org/10.1016/j.physa.2022.127460},
url = {https://www.sciencedirect.com/science/article/pii/S0378437122003363},
author = {Huiqin Zhang and Jiasen Jin}
}

@Article{ground_state_schiro_scipost_2025,
	title={{Symmetries, conservation laws and entanglement in non-Hermitian fermionic lattices}},
	author={Rafael D. Soares and Youenn Le Gal and Chun Y. Leung and Dganit Meidan and Alessandro Romito and Marco Schirò},
	journal={SciPost Phys.},
	volume={19},
	pages={094},
	year={2025},
	publisher={SciPost},
	doi={10.21468/SciPostPhys.19.4.094},
	url={https://scipost.org/10.21468/SciPostPhys.19.4.094},
}

@article{turkeshi_schrio_stochastic_resetting_2022,
  title = {Entanglement transitions from stochastic resetting of non-Hermitian quasiparticles},
  author = {Turkeshi, Xhek and Dalmonte, Marcello and Fazio, Rosario and Schir\`o, Marco},
  journal = {Phys. Rev. B},
  volume = {105},
  issue = {24},
  pages = {L241114},
  numpages = {6},
  year = {2022},
  month = {Jun},
  publisher = {American Physical Society},
  doi = {10.1103/PhysRevB.105.L241114},
  url = {https://link.aps.org/doi/10.1103/PhysRevB.105.L241114}
}

@article{sudipto_soumik_non_hermitian_eth_prl_2025,
  title = {Unveiling Eigenstate Thermalization for Non-Hermitian systems},
  author = {Singha Roy, Sudipto and Bandyopadhyay, Soumik and Costa de Almeida, Ricardo and Hauke, Philipp},
  journal = {Phys. Rev. Lett.},
  volume = {134},
  issue = {18},
  pages = {180405},
  numpages = {7},
  year = {2025},
  month = {May},
  publisher = {American Physical Society},
  doi = {10.1103/PhysRevLett.134.180405},
  url = {https://link.aps.org/doi/10.1103/PhysRevLett.134.180405}
}

@article{kevin_hauke_prxq_2022,
  title = {From Non-Hermitian Linear Response to Dynamical Correlations and Fluctuation-Dissipation Relations in Quantum Many-Body Systems},
  author = {Geier, Kevin T. and Hauke, Philipp},
  journal = {PRX Quantum},
  volume = {3},
  issue = {3},
  pages = {030308},
  numpages = {34},
  year = {2022},
  month = {Jul},
  publisher = {American Physical Society},
  doi = {10.1103/PRXQuantum.3.030308},
  url = {https://link.aps.org/doi/10.1103/PRXQuantum.3.030308}
}

@footnote{matteo_hauke_non_herm,
note={Non-Hermitian Hamiltonian engineering in qudit systems,  M. M. Wauters, P. Boschetto, E. Ballini, A. Biella and P. Hauke, will be available in arXiv soon.}
}

@article{giorgi_dissipative_resource_quantum_reservoir_engineering,
  doi = {10.22331/q-2024-03-20-1291},
  url = {https://doi.org/10.22331/q-2024-03-20-1291},
  title = {Dissipation as a resource for {Q}uantum {R}eservoir {C}omputing},
  author = {Sannia, Antonio and Mart{\'{i}}nez-Pe{\~{n}}a, Rodrigo and Soriano, Miguel C. and Giorgi, Gian Luca and Zambrini, Roberta},
  journal = {{Quantum}},
  issn = {2521-327X},
  publisher = {{Verein zur F{\"{o}}rderung des Open Access Publizierens in den Quantenwissenschaften}},
  volume = {8},
  pages = {1291},
  month = mar,
  year = {2024}
}

@article{turkeshi_no_click_limit_2021,
  title = {Measurement-induced entanglement transitions in the quantum Ising chain: From infinite to zero clicks},
  author = {Turkeshi, Xhek and Biella, Alberto and Fazio, Rosario and Dalmonte, Marcello and Schir\'o, Marco},
  journal = {Phys. Rev. B},
  volume = {103},
  issue = {22},
  pages = {224210},
  numpages = {13},
  year = {2021},
  month = {Jun},
  publisher = {American Physical Society},
  doi = {10.1103/PhysRevB.103.224210},
  url = {https://link.aps.org/doi/10.1103/PhysRevB.103.224210}
}

@article{Daley_review_quantum_jumps,
author = {Andrew J. Daley},
title = {Quantum trajectories and open many-body quantum systems},
journal = {Advances in Physics},
volume = {63},
number = {2},
pages = {77--149},
year = {2014},
publisher = {Taylor \& Francis},
doi = {10.1080/00018732.2014.933502},


URL = { 
    
        https://doi.org/10.1080/00018732.2014.933502
    
    

},
eprint = { 
    
        https://doi.org/10.1080/00018732.2014.933502
    
    

}

}

@article{quantum_jump_plenio,
  title = {The quantum-jump approach to dissipative dynamics in quantum optics},
  author = {Plenio, M. B. and Knight, P. L.},
  journal = {Rev. Mod. Phys.},
  volume = {70},
  issue = {1},
  pages = {101--144},
  numpages = {0},
  year = {1998},
  month = {Jan},
  publisher = {American Physical Society},
  doi = {10.1103/RevModPhys.70.101},
  url = {https://link.aps.org/doi/10.1103/RevModPhys.70.101}
}

@misc{geier2021noninvasivemeasurementcurrentsanalog,
      title={Non-invasive measurement of currents in analog quantum simulators}, 
      author={Kevin T. Geier and Janika Reichstetter and Philipp Hauke},
      year={2021},
      eprint={2106.12599},
      archivePrefix={arXiv},
      primaryClass={quant-ph},
      url={https://arxiv.org/abs/2106.12599}, 
}

@article{zoller_reservoir_engineering_1996,
  title = {Quantum Reservoir Engineering with Laser Cooled Trapped Ions},
  author = {Poyatos, J. F. and Cirac, J. I. and Zoller, P.},
  journal = {Phys. Rev. Lett.},
  volume = {77},
  issue = {23},
  pages = {4728--4731},
  numpages = {0},
  year = {1996},
  month = {Dec},
  publisher = {American Physical Society},
  doi = {10.1103/PhysRevLett.77.4728},
  url = {https://link.aps.org/doi/10.1103/PhysRevLett.77.4728}
}

@article{clerk_reservoir_engineering_non_reciprocal_2015,
  title = {Nonreciprocal Photon Transmission and Amplification via Reservoir Engineering},
  author = {Metelmann, A. and Clerk, A. A.},
  journal = {Phys. Rev. X},
  volume = {5},
  issue = {2},
  pages = {021025},
  numpages = {16},
  year = {2015},
  month = {Jun},
  publisher = {American Physical Society},
  doi = {10.1103/PhysRevX.5.021025},
  url = {https://link.aps.org/doi/10.1103/PhysRevX.5.021025}
}

@article{fazio_persistent_current_reservoir_engineering_2018,
  title = {Persistent currents by reservoir engineering},
  author = {Keck, Maximilian and Rossini, Davide and Fazio, Rosario},
  journal = {Phys. Rev. A},
  volume = {98},
  issue = {5},
  pages = {053812},
  numpages = {10},
  year = {2018},
  month = {Nov},
  publisher = {American Physical Society},
  doi = {10.1103/PhysRevA.98.053812},
  url = {https://link.aps.org/doi/10.1103/PhysRevA.98.053812}
}

@article{gorshkov_ground_state_hard_prl_2019,
  title = {Probing Ground-State Phase Transitions through Quench Dynamics},
  author = {Titum, Paraj and Iosue, Joseph T. and Garrison, James R. and Gorshkov, Alexey V. and Gong, Zhe-Xuan},
  journal = {Phys. Rev. Lett.},
  volume = {123},
  issue = {11},
  pages = {115701},
  numpages = {6},
  year = {2019},
  month = {Sep},
  publisher = {American Physical Society},
  doi = {10.1103/PhysRevLett.123.115701},
  url = {https://link.aps.org/doi/10.1103/PhysRevLett.123.115701}
}

@article{heralded_pt_,
  title = {Heralded Magnetism in Non-Hermitian Atomic Systems},
  author = {Lee, Tony E. and Chan, Ching-Kit},
  journal = {Phys. Rev. X},
  volume = {4},
  issue = {4},
  pages = {041001},
  numpages = {13},
  year = {2014},
  month = {Oct},
  publisher = {American Physical Society},
  doi = {10.1103/PhysRevX.4.041001},
  url = {https://link.aps.org/doi/10.1103/PhysRevX.4.041001}
}

@article{turkeshi_prb_2023,
  title = {Entanglement and correlation spreading in non-Hermitian spin chains},
  author = {Turkeshi, Xhek and Schir\'o, Marco},
  journal = {Phys. Rev. B},
  volume = {107},
  issue = {2},
  pages = {L020403},
  numpages = {6},
  year = {2023},
  month = {Jan},
  publisher = {American Physical Society},
  doi = {10.1103/PhysRevB.107.L020403},
  url = {https://link.aps.org/doi/10.1103/PhysRevB.107.L020403}
}

@article{soumik_sudipto_hauke_2025,
  title = {Quantifying non-Hermiticity using single- and many-particle quantum properties},
  author = {Bandyopadhyay, Soumik and Hauke, Philipp and Singha Roy, Sudipto},
  journal = {Phys. Rev. B},
  volume = {112},
  issue = {13},
  pages = {134201},
  numpages = {13},
  year = {2025},
  month = {Oct},
  publisher = {American Physical Society},
  doi = {10.1103/xm96-zxsw},
  url = {https://link.aps.org/doi/10.1103/xm96-zxsw}
}
\end{document}